\documentclass[twocolumn]{aastex63}

\newcommand{\lya}{{Ly$\alpha$}}

\newcommand{\hb}{H$\beta$}
\newcommand{\oiii}{[O\,{\sc iii]} $\lambda$5007}
\newcommand{\ha}{H$\alpha$}

\newcommand{\nv}{N\,{\sc v}}

\received{May 05, 2020}
\revised{Sep 04, 2020}
\accepted{Sep 08, 2020}
\submitjournal{ApJ}

\shorttitle{\lya\ Emitters at $z \approx 5.7$}
\shortauthors{Ning et al.}

\begin{document}

\title{The Magellan M2FS Spectroscopic Survey of High-Redshift Galaxies: A Sample of 260 Ly$\alpha$ Emitters at Redshift $z\approx5.7$}

\author[0000-0001-9442-1217]{Yuanhang Ning}
\affiliation{Kavli Institute for Astronomy and Astrophysics, Peking University, Beijing 100871, China}
\affiliation{Department of Astronomy, School of Physics, Peking University, Beijing 100871, China}

\author[0000-0003-4176-6486]{Linhua Jiang}
\altaffiliation{jiangKIAA@pku.edu.cn}
\affiliation{Kavli Institute for Astronomy and Astrophysics, Peking University, Beijing 100871, China}
\affiliation{Department of Astronomy, School of Physics, Peking University, Beijing 100871, China}

\author[0000-0002-9634-2923]{Zhen-Ya Zheng}
\affiliation{CAS Key Laboratory for Research in Galaxies and Cosmology, 	Shanghai Astronomical Observatory, Shanghai 200030, China}

\author[0000-0002-6168-3867]{Jin Wu}
\affiliation{Kavli Institute for Astronomy and Astrophysics, Peking University, Beijing 100871, China}
\affiliation{Department of Astronomy, School of Physics, Peking University, Beijing 100871, China}

\author[0000-0002-1620-0897]{Fuyan Bian}
\affiliation{European Southern Observatory, Alonso de C\'{o}rdova 3107, Casilla 19001, Vitacura, Santiago 19, Chile}

\author[0000-0003-1344-9475]{Eiichi Egami}
\affiliation{Steward Observatory, University of Arizona, 933 North Cherry Avenue, Tucson, AZ 85721, USA}

\author[0000-0003-3310-0131]{Xiaohui Fan}
\affiliation{Steward Observatory, University of Arizona, 933 North Cherry Avenue, Tucson, AZ 85721, USA}

\author[0000-0001-6947-5846]{Luis C. Ho}
\affiliation{Kavli Institute for Astronomy and Astrophysics, Peking University, Beijing 100871, China}
\affiliation{Department of Astronomy, School of Physics, Peking University, Beijing 100871, China}

\author[0000-0003-1659-7035]{Yue Shen}
\affiliation{Department of Astronomy, University of Illinois at Urbana-Champaign, Urbana, IL 61801, USA}
\affiliation{National Center for Supercomputing Applications, University of Illinois at Urbana-Champaign, Urbana, IL 61801, USA}

\author[0000-0003-4956-5742]{Ran Wang}
\affiliation{Kavli Institute for Astronomy and Astrophysics, Peking University, Beijing 100871, China}

\author[0000-0002-7350-6913]{Xue-Bing Wu}
\affiliation{Kavli Institute for Astronomy and Astrophysics, Peking University, Beijing 100871, China}
\affiliation{Department of Astronomy, School of Physics, Peking University, Beijing 100871, China}

\begin{abstract}

We present a spectroscopic survey of \lya\ emitters (LAEs) at $z\approx5.7$ using the multi-object spectrograph M2FS on the Magellan Clay telescope. This is part of a high-redshift galaxy survey carried out in several well-studied deep fields. These fields have deep images in multiple UV/optical bands, including a narrow NB816 band that has allowed an efficient selection of LAE candidates at $z\approx5.7$. Our sample consists of 260 LAEs and covers a total effective area of more than two square degrees on the sky. This is so far the largest (spectroscopically confirmed) sample of LAEs at this redshift. We use the secure redshifts and narrowband photometry to measure \lya\ luminosities. We find that these LAEs span a \lya\ luminosity range of $\sim 2\times10^{42} - 5\times10^{43}$ erg s$^{-1}$, and include some of the most luminous galaxies known at $z \ge 5.7$ in terms of \lya\ luminosity. Most of them have rest-frame equivalent widths between 20 and 300 \AA, and more luminous \lya\ emission lines tend to have broader line widths. We detect a clear offset of $\sim20$ \AA\ between the observed \lya\ wavelength distribution and the NB816 filter transmission curve, which can be explained by the intergalactic medium absorption of continua blueward of \lya\ in the high-redshift spectra. This sample is being used to study the \lya\ luminosity function and galaxy properties at $z\approx5.7$.

\end{abstract}


\section{Introduction}

The \lya\ emission line was predicted as a prominent feature in the spectra of early-stage galaxies \citep{pp67}. It is a powerful tracer to discover and study young star-forming galaxies at high redshift. Now \lya\ emitters (LAEs) at redshifts up to $z\ge6$ are being routinely found \citep[e.g.,][]{rhoads00, ellis01, hu10, ouchi10, kashikawa11, erb14, zheng16, tilvi20}. These LAEs can help us understand not only the evolution and physics of high-redshift galaxies \citep[e.g.,][]{finkelstein12, bouwens14, cl16, jiang16a}, but also the epoch of cosmic reionization \citep[e.g.,][]{malhotra04, kashikawa06, kashikawa11, hu10, pentericci14, santos16, ota17}.

High-redshift LAEs are usually selected in narrowband imaging surveys. The narrowband technique can efficiently detect \lya\ emission lines in LAEs. Three optical atmospheric windows with little OH sky emission have often been used to find LAEs at redshift slices around 5.7, 6.5, 7.0 \citep[e.g.,][]{taniguchi05, kashikawa06, kashikawa11, hu10, ouchi10, rhoads12, matthee15, konno18}. In addition, a number of LAEs have been spectroscopically confirmed at these redshifts \citep[e.g.,][]{hu17, jiang17, matthee17, zheng17, shibuya18b, taylor20}. Narrowband-selected LAEs at $z>7$ have also been reported \citep[e.g.,][]{tilvi10, shibuya12}.

Despite the progress that has been made so far, the number of spectroscopically confirmed LAEs at $z\ge5.7$ is still relatively small. Most LAEs were from photometrically selected samples in wide-field narrowband surveys. Some studies have covered more than 10 deg$^2$, and most of them only targeted the most luminous LAEs \citep[e.g.,][]{matthee15, hu16, santos16, konno18, shibuya18a, taylor20}. There exist large discrepancies in measurements of \lya\ luminosity functions (LFs) between photometrically selected samples and spectroscopically confirmed samples \citep[e.g.,][]{matthee15, santos16, bagley17}. There are also significant discrepancies in the \lya\ LF measurements among different spectroscopic samples \citep[e.g.,][]{kashikawa06, kashikawa11, hu10, ouchi10}. The reason for these discrepancies is not clear, and it may include sample contamination and cosmic variance. Therefore, we need a much larger LAE sample with spectroscopic redshifts over a large sky area.

In this paper, we present a spectroscopic sample of 260 LAEs at $z\approx5.7$ in five well-studied deep fields. This is part of our spectroscopic survey of high-redshift galaxies using the multi-object spectrograph, the Michigan/{\it Magellan} Fiber System (M2FS), on the 6.5m Magellan Clay telescope. We aim to build large samples of galaxies including LAEs at $z\approx5.7$ and 6.5 and Lyman-break galaxies (LBGs) at $5.5<z<6.8$. The program overview paper provides more details \citep{jiang17}. Using this LAE sample, we have detected diffuse \lya\ halos around $z\approx5.7$ LAEs \citep{wu20}. We have also discovered a giant protocluster at $z\approx5.7$ \citep{jiang18}. In this paper we will provide the details of the $z\approx5.7$ LAE sample and release the galaxy catalog. We will present the \lya\ LF of the LAEs at $z\approx5.7$ in a following paper.

The paper has a layout as follows. In Section 2, we introduce the M2FS survey program, our target selection,  spectroscopic observations, and data reduction. In Section 3, we identify LAEs and contaminants, and construct our LAE sample. In Section 4, we measure the \lya\ spectral properties of the $z\approx5.7$ LAEs in our sample. We discuss our results in Section 5 and summarize our paper in Section 6. We provide the detailed information of the full sample, including their one-dimensional (1D) and two-dimensional (2D) spectra. Throughout the paper, we use a standard flat cosmology with $H_0=\rm{68\ km\ s^{-1}\ Mpc^{-1}}$, $\Omega_m=0.3$ and $\Omega_{\Lambda}=0.7$. All magnitudes refer to the AB system.

\floattable
\begin{deluxetable}{ccccccccccc}
\tablecaption{Survey Fields
\label{fieldsinfo}}
\tablewidth{0pt}
\tablehead{
\colhead{Field} & \colhead{Coordinates} & \colhead{Area} & \colhead{Filters} & \colhead{$R/r'$} & \colhead{$I/i'$} & \colhead{$z'$} & \colhead{NB816} & \colhead{Candidates} & \colhead{Targets} & \colhead{Confirmed}\\
\colhead{} & \colhead{(J2000.0)} & \colhead{(deg$^2$)} & \colhead{} & \colhead{(mag)} & \colhead{(mag)} & \colhead{(mag)} & \colhead{(mag)} & \colhead{} & \colhead{} & \colhead{}
}
\colnumbers
\startdata
	SXDS	&	02:18:00 --05:00:00		&	1.12	&	$R\ i'\ z'$&	27.4	&	27.4	&	26.2	&	26.1	&	263 (99)	&	185 (74)	&	130 (44)	\\
	A370a	&	02:39:55 --01:35:24		&	0.16	&	$R\ I\ z'$&	27.0	&	26.2	&	26.3	&	26.0	&	\ 75 (30)	&	\ 68 (28)	&	\ 52 (18)	\\
	ECDFS	&	03:32:25 --27:48:18		&	0.22	&	$r'\ i'\ z'$&	27.4	&	27.5	&	26.7	&	26.0	&	\ 27 (11)	&	18 (9)	&	11 (5)		\\
	COSMOS	&	10:00:29 +02:12:21		&	1.26	&	$r'\ i'\ z'$&	26.7	&	26.3	&	25.5	&	25.7	&\ 228 (140)	&	158 (93)	&	\ 52 (15)	\\
	SSA22a	&	22:17:32 +00:15:14		&	0.17	&	$R\ I\ z'$&	28.0	&	27.3	&	26.7	&	26.1	&	23 (5)	&	20 (4)	&	15 (3)	\\
\enddata
\tablecomments{Columns 5, 6, 7, and 8 indicate the magnitude limits ($5\sigma$ detections in a $2\arcsec$-diameter aperture). Column 9 indicates the total number of  LAE candidates at $z\approx5.7$ in each field. Column 10 indicates the number of candidates  observed by our M2FS program. Column 11 indicates the number of the confirmed LAEs. The numbers in parenthesis represent the sources selected by our relaxed criteria (see details in section 2.2).}
\end{deluxetable}

\section{Target Selection and Spectroscopic Observations}

In this section, we will first provide a brief review of our Magellan M2FS spectroscopic survey of high-redshift galaxies, and present the selection of the LAE candidates at $z\approx5.7$ in detail. We will then outline the M2FS observations of the candidates. We will also present our data reduction pipeline which has been slightly improved from the previous version.

\subsection{The Magellan M2FS Survey}

Our M2FS survey is a spectroscopic survey of galaxies at $z>5.5$ using Magellan M2FS. M2FS is a fiber-fed, multi-object, double optical spectrograph on the Magellan Clay telescope \citep{mateo12}. The survey aims to build a large and homogeneous sample of relatively luminous LAEs at $z\approx$ 5.7 and 6.5, and LBGs with strong \lya\ emission at $5.5<z<6.8$. The target candidates come from five well-studied deep fields, including the Subaru {\it XMM-Newton} Deep Survey (SXDS), A370, the Extended {\it Chandra} Deep Field-South (ECDFS), COSMOS, and SSA22, covering a sky area of $>3$ deg$^2$ in total. These fields have a large number of archival UV/optical images in a series of broad $[BVR(r')I(i')z']$ and narrow bands (e.g., NB816 and NB921) from Subaru Suprime-Cam. They can be used to efficiently select high-redshift LAEs and LBGs. The fields are summarized in Table \ref{fieldsinfo}. Columns 5-8 list the magnitude limits of the broadband and NB816-band images. The average depth ($5\sigma$ detections in a $2\arcsec$-diameter aperture) is $\sim27.0$ mag in $R/r'$ and $I/i'$, $\sim26.5$ mag in $z'$, and $\sim26.0$ mag in NB816. Our program overview paper \citep{jiang17} provides more details about the survey program, including the survey fields, imaging data, spectroscopic observations, data reduction, and science goals.

The M2FS observations of the program have been completed and the data have been reduced. The program will provide large samples of high-redshift LAEs and LBGs over more than two deg$^2$. This will enable many science goals, such as the \lya\ luminosity function and its evolution at high redshift, properties of LAEs and LBGs, high-redshift protoclusters, cosmic reionization, etc. In this paper, we focus on LAEs at $z\approx5.7$.

\subsection{Candidate Selection}

\begin{figure}[t]
\epsscale{1.15}
\plotone{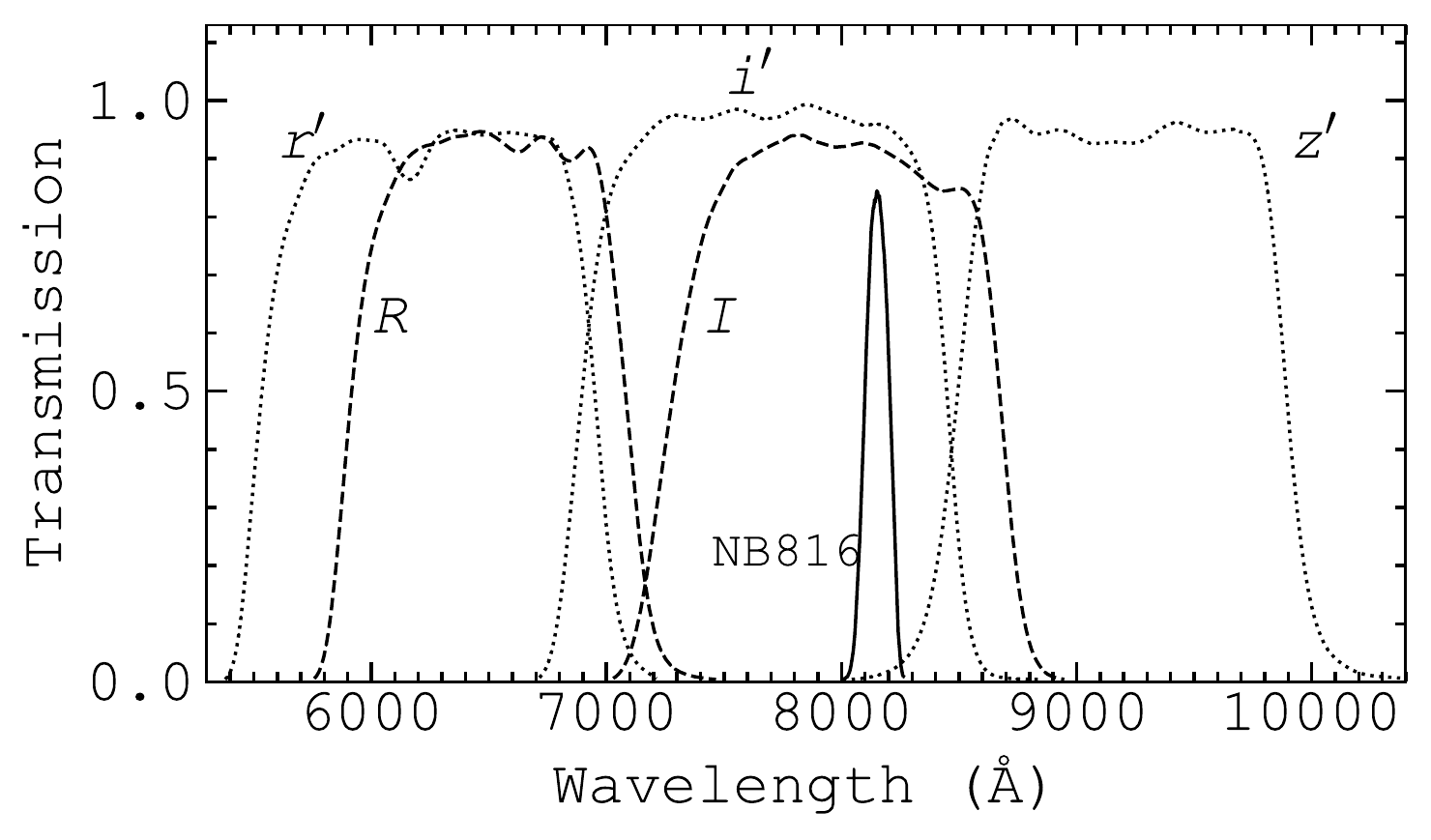}
\caption{Transmission curves of the Suprime-Cam filters that were used for 
our target selection. The NB816 band corresponds to the detection of LAEs 
at $z\approx5.7$.
\label{filters}}
\end{figure}

In the literature, LAE candidates are usually selected by the narrowband (or \lya) technique. Figure \ref{filters} shows the filters that we used for our target selection. We mainly used the $i-{\rm NB816}$ color to select $z\approx5.7$ LAE candidates (here $i$ means either $i'$ or $I$). Different fields have slightly different combinations of the broadband filters, such as $r'i'z'$, $Ri'z'$, and $RIz'$ (see Column 4 in Table \ref{fieldsinfo}).

For all $>7\sigma$ detections in the NB816 band, we applied the following color cut, 
\begin{eqnarray}
   I-{\rm NB816}>0.8
\end{eqnarray}
for the A370 and SSA22, and
\begin{eqnarray}
   i'-{\rm NB816}>1.0
\end{eqnarray}
for the SXDS, ECDFS and COSMOS fields. The target selection (color-magnitude diagram) of the $z\approx5.7$ candidates in SXDS is illustrated in Figure 7 of \citet{jiang17}. The $i'$ filter is slightly bluer than the $I$ filter, and thus suffers more Lyman forest absorption blueward the \lya\ emission line. Therefore, we used the slightly different criteria to ensure a similar broadband -- narrowband color. The two criteria are similar to those used in the literature \citep[e.g.,][]{ouchi08, hu10}, and roughly correspond to a \lya\ rest-frame equivalent width (EW) limit of $\sim25$ \AA. Since the $i'$ and $I$-band images are much deeper than the NB816-band images, objects undetected in $i'$ or $I$ naturally satisfy the color selection.

We applied two additional criteria to eliminate lower-redshift contaminants. We required that candidates should not be detected ($<2\sigma$) in any band (e.g., $B$ or $V$) bluer than $R$ or $r'$, assuming that no flux can be detected at a wavelength bluer than the Lyman limit. We also applied a color selection of $r' (R)-z' > 1.5$ for objects detected (at $>3\sigma$) in $z'$. These two criteria do not remove real $z\sim5.7$ objects. Each candidate was visually inspected. We removed spurious detections such as the residuals of bright star spikes and satellite trails that can be easily identified. We also removed objects whose photometry was obviously wrong due to the existence of nearby bright stars.

\floattable
\begin{deluxetable}{ccclc}
\tablecaption{Summary of the M2FS Observations
\label{m2fsobs}}
\tablehead{
\colhead{Pointing} & \colhead{Center Coordinates} & \colhead{Field Coverage} 
& \colhead{Year/Month} & \colhead{Exposure Time}  \\
\colhead{} & \colhead{(J2000.0)} & \colhead{(deg$^2$)} & 
\colhead{} & \colhead{(hours)} }
\colnumbers
\startdata
SXDS1		& 02:18:18.2 --05:00:09.96	& 0.179	&  2016 Dec	&  4.0	\\
			& 						& 		&  2017 Sep	&  2.0	\\
SXDS2		& 02:17:47.8 --04:35:26.63	& 0.185	&  2016 Dec	&  5.0	\\
SXDS3		& 02:17:46.0 --05:26:17.88	& 0.182	&  2015 Nov	&  7.0	\\
SXDS4		& 02:19:43.5 --05:01:39.25	& 0.187	&  2018 Dec	&  6.8	\\
SXDS5		& 02:16:16.6 --05:00:45.04	& 0.188	&  2016 Dec	&  5.0	\\
A370a		& 02:39:49.4 --01:35:12.16	& 0.173	&  2015 Sep	&  7.0	\\
ECDFS		& 03:31:59.8 --27:49:17.07	& 0.145	&  2016 Feb	&  6.3	\\
COSMOS1		& 10:01:45.4 +02:23:43.76	& 0.197    &  2015 Apr	&  4.0	\\
COSMOS2	& 09:59:59.3 +02:26:30.96	& 0.197	&  2015 Apr	&  4.5	\\
COSMOS3 	& 10:01:28.3 +01:59:36.21 	& 0.197	&  2015 Apr	&  5.0	\\
COSMOS4	& 09:59:32.4 +02:00:33.39	& 0.197	&  2015 Apr	&  5.0	\\
COSMOS5 	& 09:59:18.3 +01:43:01.99		& 0.127	&  2016 Feb	&  5.7	\\
SSA22a		& 22:17:26.5 +00:13:40.89	& 0.171	&  2018 May	&  2.0	\\
			& 						& 		&  2018 Aug	&  4.8	\\
\enddata
\end{deluxetable}

In addition to the above main candidates, we also included a small number of less promising or fainter candidates to fill spare fibers. For example, we observed many LAEs with $\sim5\sigma-7\sigma$ detections in NB816. We summarize our candidate selection in Table \ref{fieldsinfo}, including these less promising candidates (numbers in parenthesis). From these additional sources, we identified 85 LAEs in total. These LAEs are less complete compared to the main sample. In the table, ``candidates" represent the sources selected by the color-magnitude criteria and ``targets" represent those observed by M2FS (see also Figure \ref{fields}).

\subsection{Spectroscopic Observations}
We used M2FS to carry out spectroscopic observations in 2015--2018. M2FS has a large field-of-view of $30'$ in diameter and high throughput. It can efficiently detect relatively bright, high-redshift galaxies. We used a pair of red-sensitive gratings with a resolving power of about 2000. The wavelength coverage was roughly from 7600 to 9600 \AA. We binned pixels ($2\times2$) during our observations, and the spectral dispersion was $\sim1$ \AA\ per pixel.

The selection of M2FS pointing centers was limited by the number and spatial distribution of bright stars in each field. Each field (plate or pointing) needs a Shack-Hartmann star ($V\leq14\rm\ mag$) in the center, two or more guide stars ($V\leq15\rm\ mag$), and four to eight alignment stars ($V\leq15.5\rm\ mag$). Some candidates were not covered by the M2FS pointings \citep[see][Figure 1--5]{jiang17}. In the end, more than 70\% (449 out of 616) of the $z\approx5.7$ LAE candidates were observed by 13 M2FS pointings. In addition, each pointing also covered $z\approx6.5$ LAE candidates, $z\approx6$ LBG candidates, a variety of ancillary targets, several bright reference stars, and a few tens (typically around 50) of sky fibers.

The information about the M2FS observations is summarized in Table \ref{m2fsobs}. Column 1 shows the M2FS pointing or field names. SXDS1, SXDS2, SXDS3, SXDS4, and SXDS5 denote the five pointings in SXDS. COSMOS1, COSMOS2, COSMOS3, COSMOS4, and COSMOS5 denote the five pointings in COSMOS. The layout of the pointings are shown later in Figure \ref{fields}. All M2FS observations are carried out in queue mode. COSMOS1 and COSMOS3 are the first two fields that we observed. After the observations of the two fields, we checked the spectra of bright references stars and noticed that the two fields suffered serious alignment problems. The reason is unclear. The consequence is that we only confirmed a few LAEs in the two fields (see section 3).
 
Most data ($>90$\%) were taken under clear observing conditions with seeing around $0\farcs7 - 1\farcs0$. They are the data that we will use later (also shown in Column 5 of Table \ref{m2fsobs}). Data taken under cloudy weather conditions ($\sim8$\%) were not used. The effective integration time per pointing was about 5 hrs on average. The individual exposure time was 30 min, 45 min, or 1 hr, depending on airmass and weather conditions. We achieved our goal and detected $z\approx5.7$ LAEs down to $\rm NB816 \sim 25.7$ mag (a \lya\ flux depth of $\sim0.8 \times 10^{-17}$ erg s$^{-1}$ cm$^{-2}$) on average.

\subsection{Data Reduction}

\begin{figure}[t]
\centering
\plotone{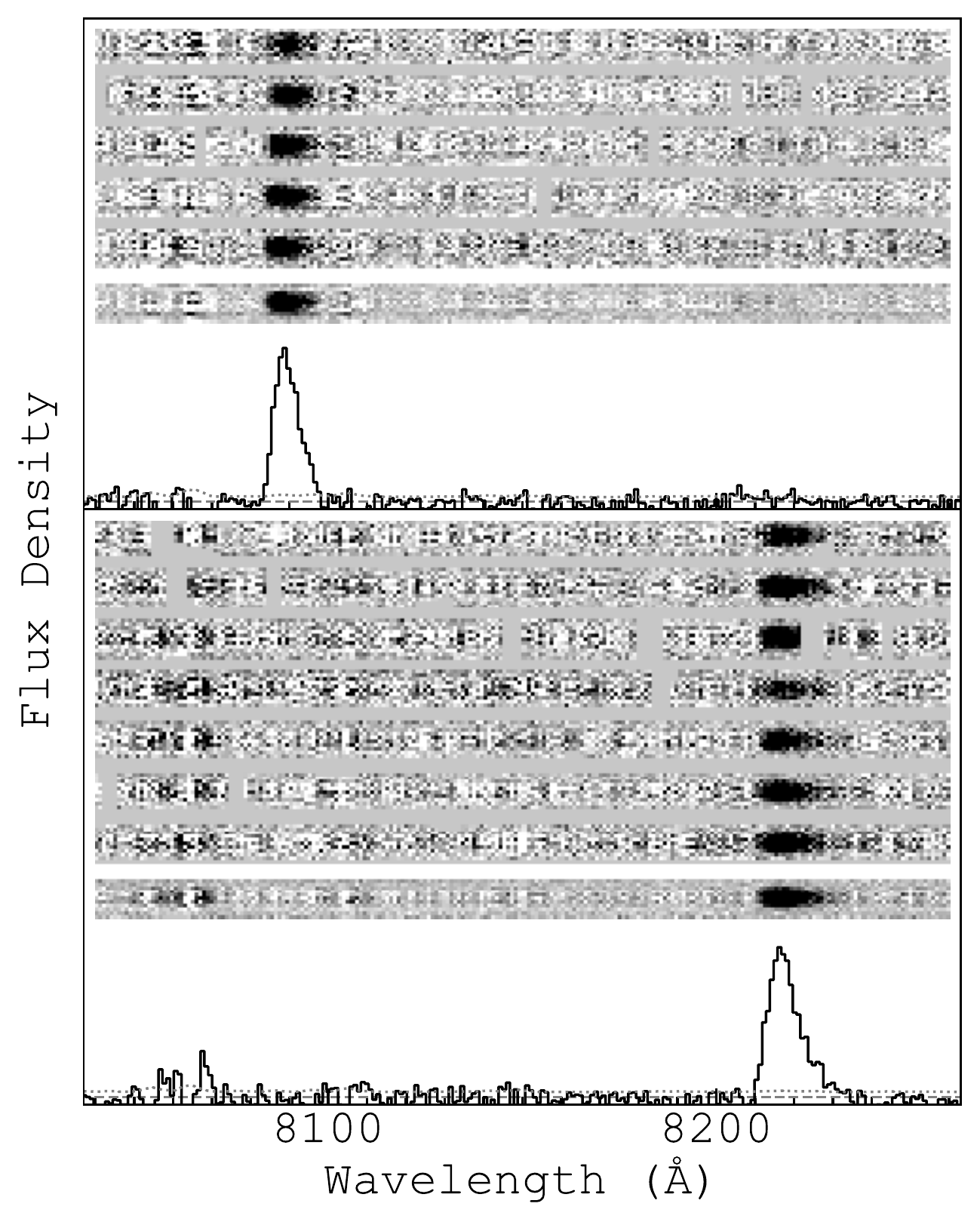}
\caption{1D and 2D spectra of two bright LAEs at $z\approx5.7$.
For each object in the upper and lower panels, we show from the top to the bottom, the 2D spectra of the individual exposures, the combined 2D spectrum, and the combined 1D spectrum.
\label{blae2max}}
\end{figure}

Our customized pipeline for data reduction is introduced in the program overview paper. The basic procedure has not been changed. This includes bias (overscan) correction, dark subtraction, flat-fielding, cosmic ray identification, and production of ``calibrated" 2D images. After fiber positions are traced using twilight images, 1D spectra are extracted from science, twilight, and lamp images. The wavelength solutions are derived from the 1D lamp spectra. For each 1D science spectrum, a sky spectrum is built by averaging the nearest $\sim10-20$ sky fibers and subtracted from the science spectrum. We also produce 2D calibrated and sky-subtracted science spectra for visual inspection, based on the method in \citet{jiang17}.

We have slightly improved the pipeline by adding additional steps. A preliminary wavelength solution is measured from the 1D lamp spectrum. In rare cases that 1D lamp spectra do not have high enough signal-to-noise ratios (S/N), the improved pipeline can use strong OH skylines in science spectra for wavelength calibration. In the COSMOS field, the pointings slightly overlap, as a result of which some objects were observed twice. These objects were reduced separately for individual pointings, and then their spectra were combined (weighted average) by the improved pipeline. As we mentioned above, we produced 2D spectra for visual inspection. The pipeline can now reserve the 2D spectra of individual exposures for visual inspection. Figure \ref{blae2max} shows 1D and 2D spectra of two bright $z\approx5.7$ LAEs in SXDS. In the upper panel from the top to the bottom, we show the 2D spectra of 5 individual exposures, the combined 2D spectrum, and the combined 1D spectrum for one LAE. In the lower panel we show the other LAE that has 7 individual exposures. The individual exposure time was 1 hr. These two LAEs can be easily confirmed even in individual exposures, due to their strong \lya\ emission and the asymmetric line shape. Most of other LAEs are much fainter, and can only be identified in their combined 1D and 2D spectra.

\section{A Sample of 260 LAEs at $z \approx 5.7$}

In this section, we will identify LAEs and present our sample of 260 spectroscopically confirmed LAEs.

\subsection{LAE Identification}

\begin{figure}[t]
\epsscale{1.15}
\centering
\plotone{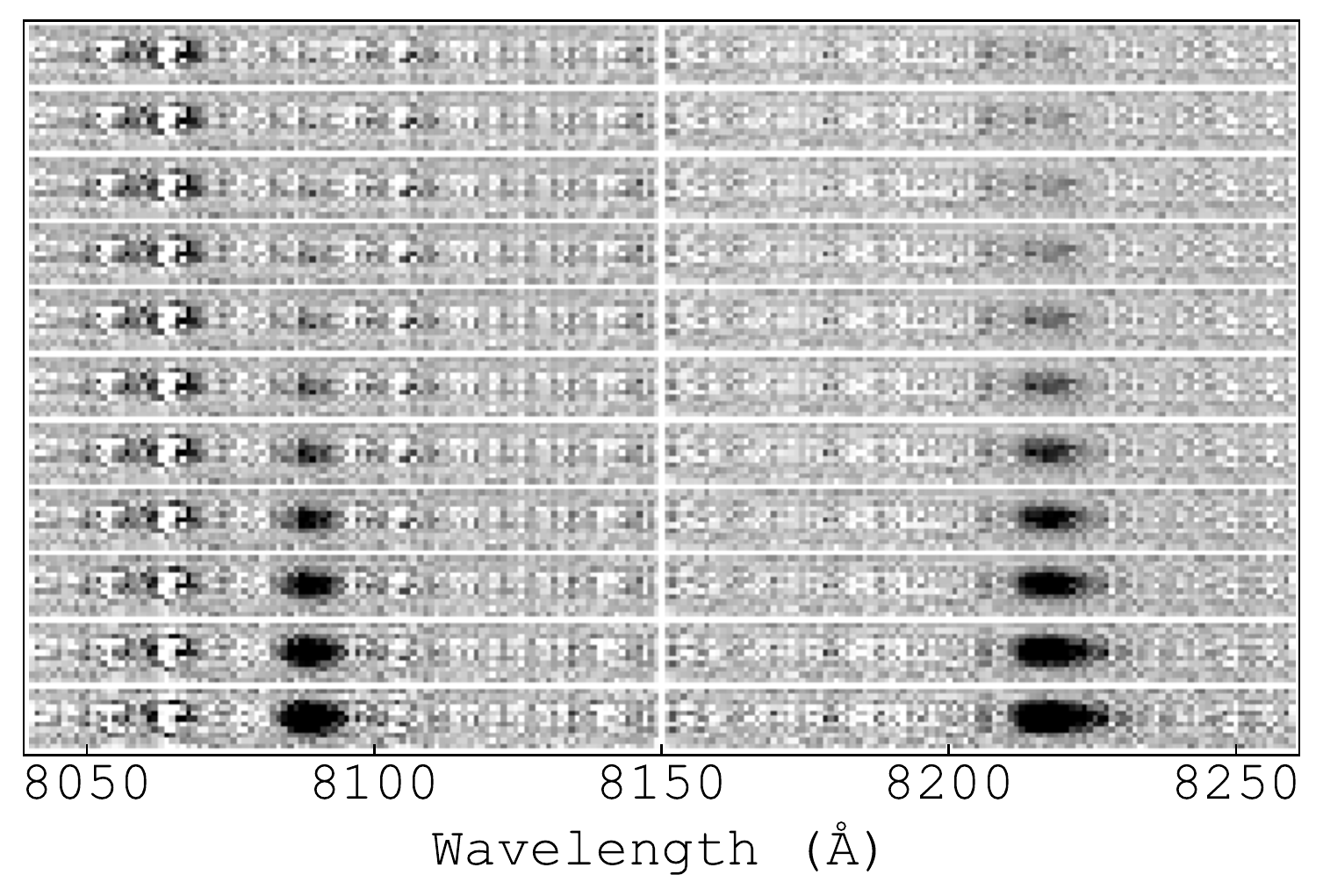}
\caption{The appearance of the 2D spectra of two \lya\ lines with decreasing S/N from the bottom to the top. The two bright LAEs in Figure \ref{blae2max} are used. See the text for details. The S/N values of the left LAE gradually decreases from $\sim$43.7 to $\sim$1.3. The S/N values of the right LAE gradually decreases from $\sim$91.7 to $\sim$2.8.
\label{blaefainter}}
\end{figure}

We use both 1D and 2D spectra to identify \lya\ emission lines. For each 1D spectrum, we first smooth it with a Gaussian kernel (a sigma of one pixel is used). We then search for an emission line with S/N $\ge5$ in the expected wavelength range. A line needs to cover at least five contiguous pixels with S/N $>1$ in the smoothed spectrum. The S/N of the line is estimated by stacking the corresponding pixels in the original spectrum. Our target selection criteria generally ensure that an emission line detected in the expected wavelength range is the \lya\ line, based on the non-detection in the deep {\it BVR} images. Next, we visually inspect the identified emission lines in the individual and combined 2D spectra.

The \lya\ emission line of a high-redshift LAE usually shows an asymmetric profile due to strong intergalactic medium (IGM) absorption and internal interstellar medium (ISM) kinematics (see Figure \ref{blae2max}). In Figure \ref{blaefainter}, we use the two bright LAEs in Figure \ref{blae2max} to illustrate how \lya\ emission lines with different S/N look like in our 2D spectra. The \lya\ lines of the two LAEs are located in two very different wavelengths that have little OH skylines. For either LAE, we first cut out a small region from its 2D spectrum that contains its \lya\ line. This region is completely dominated by the bright \lya\ line, so we assume that it is noiseless. We then scale this region by reducing the flux by a factor of $\sqrt2$ for each step. Finally, the scaled region is put onto the 2D spectrum of the other LAE. The new location of the \lya\ line is at the same wavelength where there are few skylines (i.e., this clean region is used as a true background). Each LAE is scaled 10 times. The results are shown in Figure \ref{blaefainter}. This figure shows how the appearance of a \lya\ line changes as its S/N decreases. This  serves as a reference to check LAEs in the 2D spectra. We also see that the asymmetric shape is not obvious when S/N is low. 

Based on the individual and combined 2D spectra, we can easily remove spurious or unreliable detections such as a line detection that only shows up in one of the individual 2D images, a line detection that is part of cosmic ray residuals in the 2D image, a relatively weak line that is severely contaminated by OH lines. These cases are rare. In rarer cases where a strong line is from a low-redshift galaxy, it usually appears narrow and symmetric. We will discuss this in the next subsection. 

We perform an additional test to estimate the probability of detecting a random line. We choose 100 spectra of LAE candidates at $z\approx6.5$ and search for strong line features in the wavelength range around 8160 \AA, the same range that we used to detect $z\approx5.7$ LAEs. Since we do not expect to see emission lines in this wavelength range for these targets, any line detections could be contamination. The line search follows the same procedure as we did for $z\approx5.7$ LAEs. After we remove obvious, spurious detections mentioned earlier, we do not find a strong line with S/N $\ge5$ in the 100 spectra. Therefore, the probability of detecting a random line in our sample is negligible.

Among the remaining targets that were observed by M2FS, a small fraction of them ($\sim30$) show weak emission lines (S/N $< 5$) in the wavelength range around 8160 \AA. They do not satisfy our line identification criteria and are not included in our LAE sample. They are among the faintest targets in our candidates. The rest of the targets do not have emission features in our spectra, so we do not know what objects they are.

\begin{figure}[t]
\epsscale{1.15}
\centering
\plotone{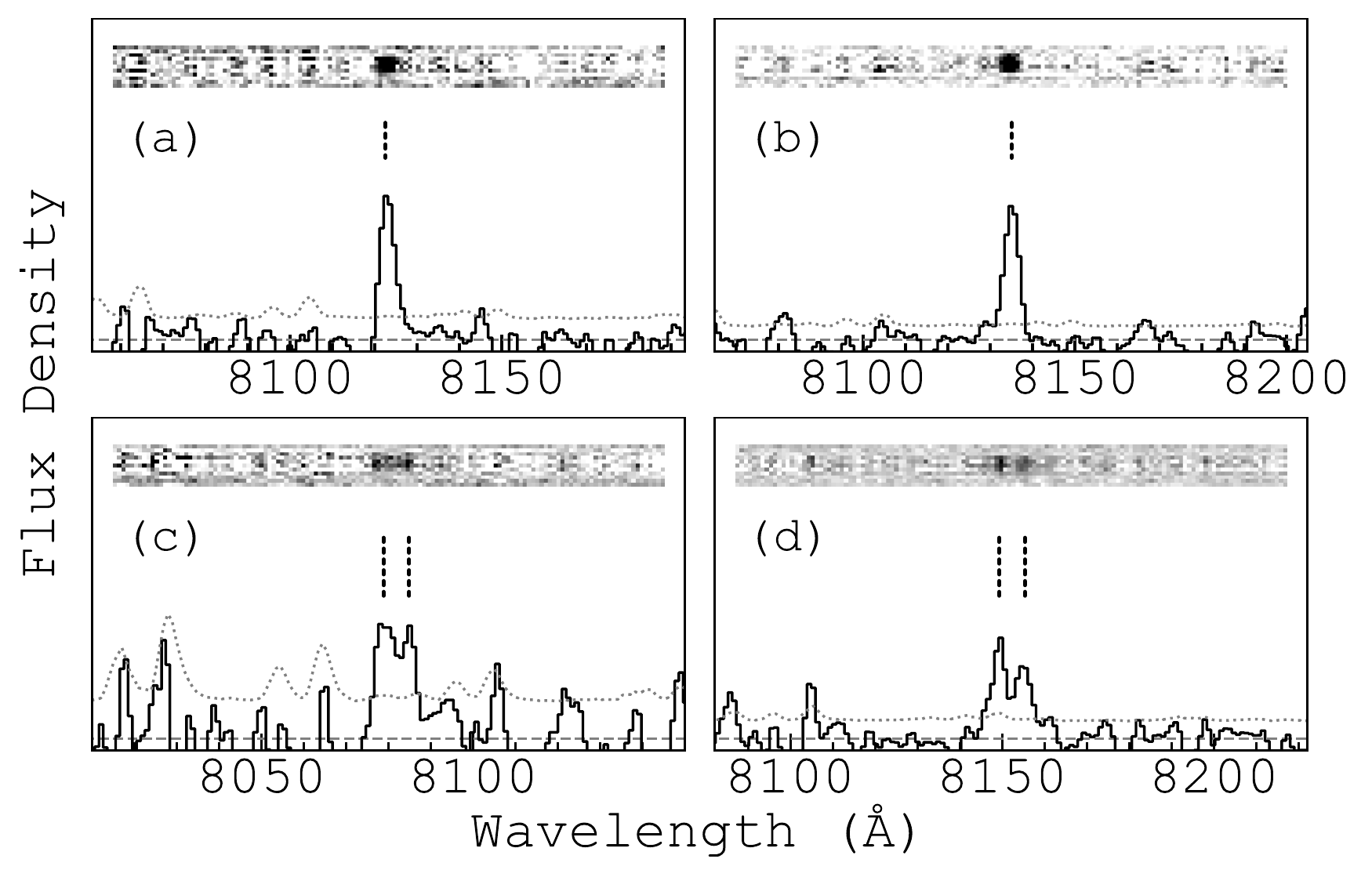}
\caption{Examples of four emission lines in our sample. The 1D spectra have been slightly smoothed with a Gaussian kernel (a $\sigma$ of one pixel is used). The dashed and dotted line indicate zero and $1\sigma$ uncertainty levels, respectively. Panel (a) represents a bright and compact line that is identified as a $z\approx5.7$ LAE, because it shows an obvious, characteristic tail in the red side of the line in both 1D and 2D spectra. Panel (b) represents a bright and compact line that is identified as a low-redshift interloper, because it shows a clear cutoff in the red side of the line in the 1D and 2D spectra. Panels (c) and (d) shows two lines that are identified as the [O\,{\sc ii}] $\lambda\lambda 3727,3729$ doublet. 
\label{plaes}}
\end{figure}

\subsection{Contaminants}

We identify and remove emission lines at $\sim8050 - 8250$ \AA\ that are likely low-redshift interlopers, including [O\,{\sc ii}] $\lambda\lambda 3727,3729$, \hb, \oiii, or \ha\ emission lines. As we mentioned earlier, our target selection criteria generally ensure that an emission line detected in the expected wavelength range is the \lya\ line. But occasionally it could be one of the above lines. The [O\,{\sc ii}] doublet is the most likely contaminant in high-redshift, narrowband-selected galaxy samples. Because there are no strong emission lines in the wavelength range between the doublet and \lya, they can be very faint in the {\it BVR} images. Our resolving power of $\sim2000$ can nearly resolve the doublet, so it is relatively easy to identify [O\,{\sc ii}]. We find five [O\,{\sc ii}] emitters in our sample. In Figure \ref{plaes}, panels (c) and (d) show two examples. In order to find possible \hb, \oiii, or \ha\ lines, we identify bright lines with compact and symmetric line shapes. We adopt the following criteria: 1) S/N $>7$; 2) line width comparable to the point spread function (PSF); 3) no obvious tail in the red side by visual inspection. We find a total of seven lines that satisfy the criteria. They are among the narrowest lines in our sample. We do not reject such lines with S/N $<7$.

In Figure \ref{plaes}, panel (a) shows a line that we identify as a real LAE at $z\approx5.7$, and panel (b) shows a line that we identify as a low-redshift galaxy. The two lines look very similar above their $1\sigma$ error lines. The real LAE clearly shows a characteristic tail on the red side of the line in both 1D (mostly below the $1\sigma$ error line) and 2D spectra, while the line in panel (b) shows a sharp cutoff on the red side of the line in its 1D and 2D spectra. It is difficult to describe such a difference quantitatively without visual inspection.
It is worth pointing out that it is likely that some of the rejected objects are real LAEs at $z\approx5.7$. For example, the object in panel (b) could be a LAE with a narrow line width and its asymmetry is not obvious. The double-peak line seen in panel (c) could be caused by its low S/N. We do not discuss more about the 12 objects. Instead, we simply remove them from our LAE sample.

\floattable
\begin{deluxetable}{ccccccccr}
\tablecaption{40 LAEs with the highest S/N in our $z\approx5.7$ sample (see the on-line table 
for the full sample)
\label{sample1}}
\tablehead{
   \colhead{No.} & \colhead{R.A.} & \colhead{Decl.} & 
   \colhead{Redshift} & \colhead{$i$} & \colhead{$z'$} & 
   \colhead{NB816} & \colhead{$L$(\lya)} & \colhead{M2FS ID}\\ 
   \colhead{} & \colhead{(J2000.0)} & \colhead{(J2000.0)} & \colhead{} & \colhead{(mag)} & 
   \colhead{(mag)} & \colhead{(mag)} & \colhead{($10^{43}\rm\ erg\ s^{-1}$)} & \colhead{}
   }
\colnumbers
\startdata
 01 &  02:40:22.35 &  $-$01:31:19.5 &    5.628 &  26.64 $\pm$ 0.39 &           $>$27.3 &  25.79 $\pm$ 0.19 &  2.04 $\pm$ 0.48 &   A370a-042 \\
 02 &  22:17:40.91 &  $+$00:24:14.5 &    5.634 &  26.50 $\pm$ 0.10 &  26.58 $\pm$ 0.19 &  25.70 $\pm$ 0.17 &  1.17 $\pm$ 0.38 &  SSA22a-019 \\
 03 &  09:58:54.51 &  $+$01:41:57.7 &    5.638 &  26.09 $\pm$ 0.11 &           $>$26.5 &  24.94 $\pm$ 0.05 &  2.58 $\pm$ 0.16 &  COSMOS-016 \\
 04 &  22:17:28.76 &  $+$00:19:17.5 &    5.643 &  26.13 $\pm$ 0.09 &  25.89 $\pm$ 0.12 &  24.86 $\pm$ 0.08 &  1.74 $\pm$ 0.24 &  SSA22a-016 \\
 05 &  02:40:29.05 &  $-$01:39:20.0 &    5.643 &  26.21 $\pm$ 0.27 &  25.85 $\pm$ 0.16 &  25.13 $\pm$ 0.11 &  1.05 $\pm$ 0.27 &   A370a-009 \\
 06 &  02:18:17.34 &  $-$05:32:23.0 &    5.644 &  26.26 $\pm$ 0.11 &  26.19 $\pm$ 0.30 &  24.58 $\pm$ 0.07 &  2.71 $\pm$ 0.31 &   SXDS3-031 \\
 07 &  02:15:55.15 &  $-$05:06:28.0 &    5.646 &  26.83 $\pm$ 0.22 &  26.84 $\pm$ 0.50 &  25.25 $\pm$ 0.18 &  1.32 $\pm$ 0.33 &   SXDS5-020 \\
 08 &  02:17:05.64 &  $-$05:32:17.7 &    5.646 &  26.20 $\pm$ 0.11 &  25.89 $\pm$ 0.23 &  25.07 $\pm$ 0.10 &  1.05 $\pm$ 0.26 &   SXDS3-033 \\
 09 &  22:17:16.37 &  $+$00:13:25.2 &    5.649 &  26.27 $\pm$ 0.11 &  25.77 $\pm$ 0.10 &  25.27 $\pm$ 0.12 &  0.55 $\pm$ 0.18 &  SSA22a-010 \\
 10 &  02:17:40.88 &  $-$04:32:36.3 &    5.653 &  26.15 $\pm$ 0.09 &  26.30 $\pm$ 0.25 &  24.06 $\pm$ 0.04 &  3.29 $\pm$ 0.17 &   SXDS2-020 \\
 11 &  02:16:05.11 &  $-$05:07:54.0 &    5.654 &  26.16 $\pm$ 0.08 &  25.23 $\pm$ 0.10 &  24.37 $\pm$ 0.06 &  1.56 $\pm$ 0.19 &   SXDS5-016 \\
 12 &  22:17:33.14 &  $+$00:22:16.0 &    5.654 &  26.04 $\pm$ 0.09 &  25.94 $\pm$ 0.12 &  25.08 $\pm$ 0.10 &  0.81 $\pm$ 0.15 &  SSA22a-018 \\
 13 &  02:39:28.58 &  $-$01:24:01.4 &    5.670 &  26.21 $\pm$ 0.29 &  26.58 $\pm$ 0.31 &  24.30 $\pm$ 0.05 &  1.65 $\pm$ 0.10 &   A370a-074 \\
 14 &  02:17:29.49 &  $-$05:38:16.6 &    5.671 &  26.20 $\pm$ 0.16 &  26.05 $\pm$ 0.35 &  24.34 $\pm$ 0.07 &  1.45 $\pm$ 0.15 &   SXDS3-001 \\
 15 &  02:20:21.51 &  $-$04:53:15.3 &    5.671 &  27.26 $\pm$ 0.26 &  26.70 $\pm$ 0.43 &  24.97 $\pm$ 0.10 &  0.81 $\pm$ 0.11 &   SXDS4-017 \\
 16 &  22:17:05.59 &  $+$00:13:00.4 &    5.671 &  26.79 $\pm$ 0.17 &  26.53 $\pm$ 0.20 &  25.02 $\pm$ 0.09 &  0.73 $\pm$ 0.08 &  SSA22a-009 \\
 17 &  02:39:17.66 &  $-$01:26:54.9 &    5.675 &  26.12 $\pm$ 0.25 &  26.03 $\pm$ 0.18 &  24.19 $\pm$ 0.05 &  1.60 $\pm$ 0.10 &   A370a-057 \\
 18 &  02:17:45.75 &  $-$04:41:29.3 &    5.676 &  27.24 $\pm$ 0.24 &           $>$27.2 &  24.82 $\pm$ 0.09 &  0.94 $\pm$ 0.09 &   SXDS2-012 \\
 19 &  02:15:59.17 &  $-$05:10:13.8 &    5.678 &  26.56 $\pm$ 0.15 &           $>$27.2 &  24.56 $\pm$ 0.08 &  1.19 $\pm$ 0.09 &   SXDS5-013 \\
 20 &  22:16:54.97 &  $+$00:05:37.9 &    5.678 &  27.51 $\pm$ 0.36 &           $>$27.7 &  24.47 $\pm$ 0.06 &  1.33 $\pm$ 0.08 &  SSA22a-003 \\
 21 &  02:17:07.87 &  $-$05:34:26.8 &    5.680 &  26.38 $\pm$ 0.13 &  26.04 $\pm$ 0.28 &  23.60 $\pm$ 0.03 &  2.77 $\pm$ 0.10 &   SXDS3-021 \\
 22 &  10:00:44.49 &  $+$02:27:19.2 &    5.684 &  27.27 $\pm$ 0.23 &           $>$26.5 &  24.97 $\pm$ 0.04 &  0.67 $\pm$ 0.03 &  COSMOS-174 \\
 23 &  09:59:05.40 &  $+$01:47:47.7 &    5.685 &           $>$27.3 &           $>$26.5 &  24.80 $\pm$ 0.04 &  0.80 $\pm$ 0.03 &  COSMOS-045 \\
 24 &  02:17:43.34 &  $-$05:28:07.1 &    5.686 &  25.95 $\pm$ 0.08 &  25.89 $\pm$ 0.23 &  23.87 $\pm$ 0.03 &  1.96 $\pm$ 0.08 &   SXDS3-116 \\
 25 &  02:17:04.30 &  $-$05:27:14.4 &    5.687 &  26.30 $\pm$ 0.11 &  26.25 $\pm$ 0.32 &  23.98 $\pm$ 0.04 &  1.80 $\pm$ 0.09 &   SXDS3-062 \\
 26 &  09:59:54.52 &  $+$02:15:16.6 &    5.689 &  26.83 $\pm$ 0.15 &           $>$26.5 &  24.66 $\pm$ 0.03 &  0.92 $\pm$ 0.03 &  COSMOS-149 \\
 27 &  02:39:53.54 &  $-$01:36:27.9 &    5.693 &  26.46 $\pm$ 0.33 &  27.13 $\pm$ 0.54 &  24.75 $\pm$ 0.07 &  0.88 $\pm$ 0.07 &   A370a-017 \\
 28 &  02:18:27.45 &  $-$04:47:37.2 &    5.703 &  26.33 $\pm$ 0.12 &  25.90 $\pm$ 0.24 &  23.86 $\pm$ 0.04 &  1.94 $\pm$ 0.09 &   SXDS1-025 \\
 29 &  02:16:24.72 &  $-$04:55:16.7 &    5.707 &  26.41 $\pm$ 0.11 &  25.91 $\pm$ 0.19 &  23.79 $\pm$ 0.04 &  2.10 $\pm$ 0.09 &   SXDS5-033 \\
 30 &  02:17:24.04 &  $-$05:33:09.7 &    5.708 &  25.67 $\pm$ 0.07 &  25.05 $\pm$ 0.11 &  23.48 $\pm$ 0.02 &  2.66 $\pm$ 0.07 &   SXDS3-029 \\
 31 &  10:00:40.22 &  $+$02:19:03.4 &    5.713 &  27.15 $\pm$ 0.20 &  26.13 $\pm$ 0.20 &  24.66 $\pm$ 0.03 &  0.92 $\pm$ 0.05 &  COSMOS-155 \\
 32 &  02:20:26.10 &  $-$04:52:35.1 &    5.720 &  25.88 $\pm$ 0.07 &  24.97 $\pm$ 0.09 &  24.25 $\pm$ 0.05 &  1.29 $\pm$ 0.08 &   SXDS4-018 \\
 33 &  03:32:37.51 &  $-$27:40:57.8 &    5.722 &  28.12 $\pm$ 0.41 &  27.53 $\pm$ 0.49 &  24.56 $\pm$ 0.06 &  1.17 $\pm$ 0.07 &   ECDFS-021 \\
 34 &  02:17:39.26 &  $-$04:38:37.4 &    5.722 &           $>$28.4 &  26.66 $\pm$ 0.36 &  25.02 $\pm$ 0.10 &  0.73 $\pm$ 0.08 &   SXDS2-014 \\
 35 &  02:39:42.90 &  $-$01:26:26.4 &    5.723 &  26.90 $\pm$ 0.52 &  26.69 $\pm$ 0.34 &  24.88 $\pm$ 0.09 &  0.84 $\pm$ 0.08 &   A370a-058 \\
 36 &  02:18:41.42 &  $-$04:52:23.0 &    5.742 &           $>$28.4 &           $>$27.2 &  25.33 $\pm$ 0.15 &  0.76 $\pm$ 0.11 &   SXDS1-020 \\
 37 &  02:39:51.27 &  $-$01:35:12.9 &    5.749 &           $>$27.2 &           $>$27.3 &  25.60 $\pm$ 0.16 &  0.72 $\pm$ 0.11 &   A370a-024 \\
 38 &  02:18:10.69 &  $-$05:37:07.8 &    5.750 &  26.82 $\pm$ 0.21 &  26.34 $\pm$ 0.36 &  25.21 $\pm$ 0.12 &  1.04 $\pm$ 0.12 &   SXDS3-007 \\
 39 &  10:00:12.79 &  $+$02:19:30.9 &    5.750 &  27.25 $\pm$ 0.23 &           $>$26.5 &  25.71 $\pm$ 0.10 &  0.65 $\pm$ 0.06 &  COSMOS-157 \\
 40 &  02:17:52.65 &  $-$05:35:11.8 &    5.759 &  25.10 $\pm$ 0.04 &  24.57 $\pm$ 0.07 &  24.04 $\pm$ 0.04 &  4.37 $\pm$ 0.17 &   SXDS3-016 \\
\enddata
\tablecomments{The upper limits listed in the table indicate $2\sigma$ detections. The redshift errors are smaller than 0.001.}
\end{deluxetable}

\begin{figure*}
\centering
\includegraphics[angle=0, width=1.\textwidth]{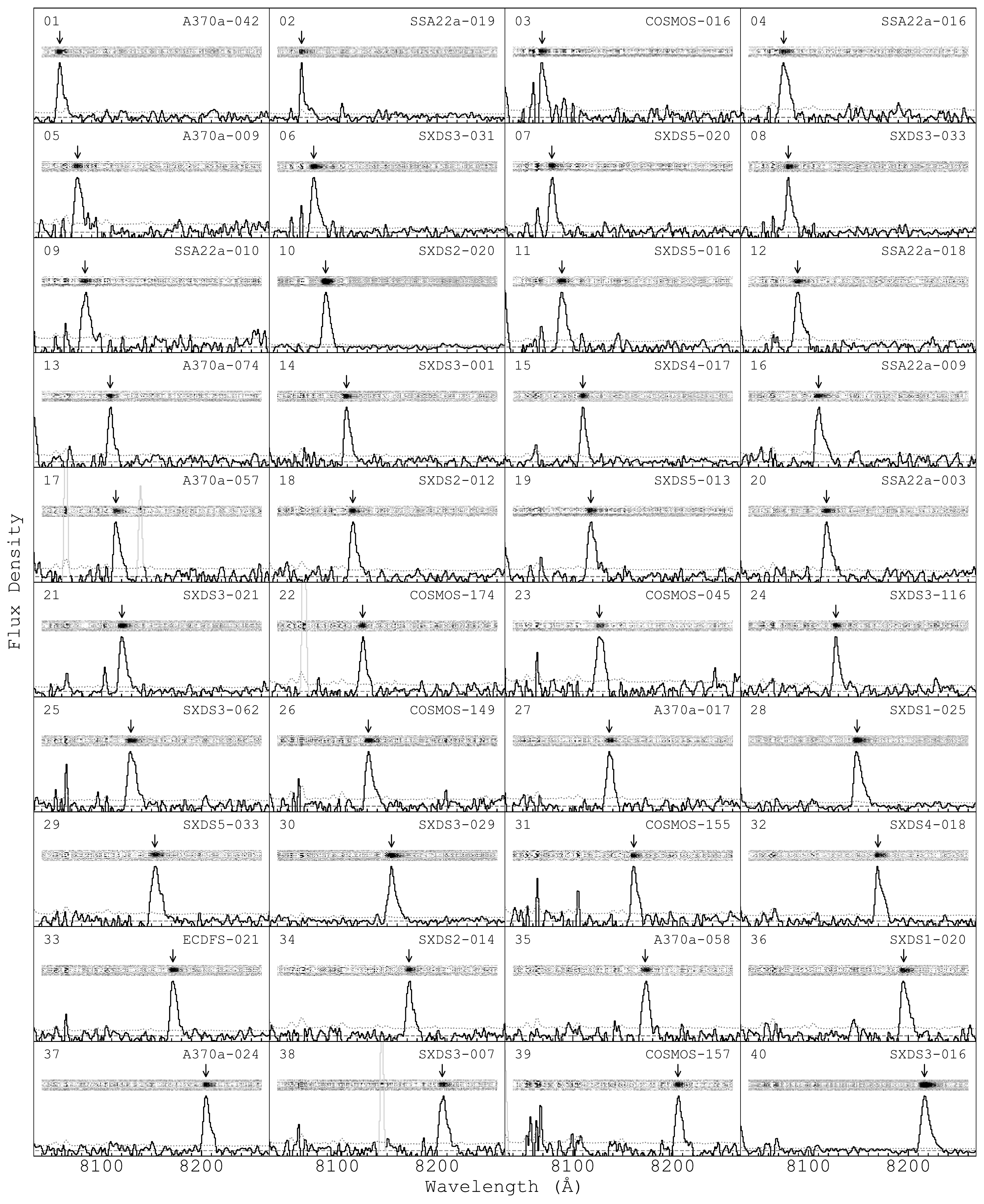}
\caption{M2FS 1D and 2D spectra of 40 LAEs with the highest S/N in our sample. The spectral dispersion is ${\sim}1$ \AA\ per pixel. The 1D spectra have been smoothed with a Gaussian kernel (a $\sigma$ of one pixel is used). In each panel, the gray dashed and dotted line indicate zero and $1\sigma$ uncertainty level, respectively. The downward arrow points to the position of the \lya\ emission line. The source number and M2FS ID correspond to those shown in Columns 1 and 9 in Table \ref{sample1}. See on-line figures for the full sample.
\label{glaeshsnr}}
\end{figure*}

\begin{figure*}
\centering
\includegraphics[angle=0, width=1.\textwidth]{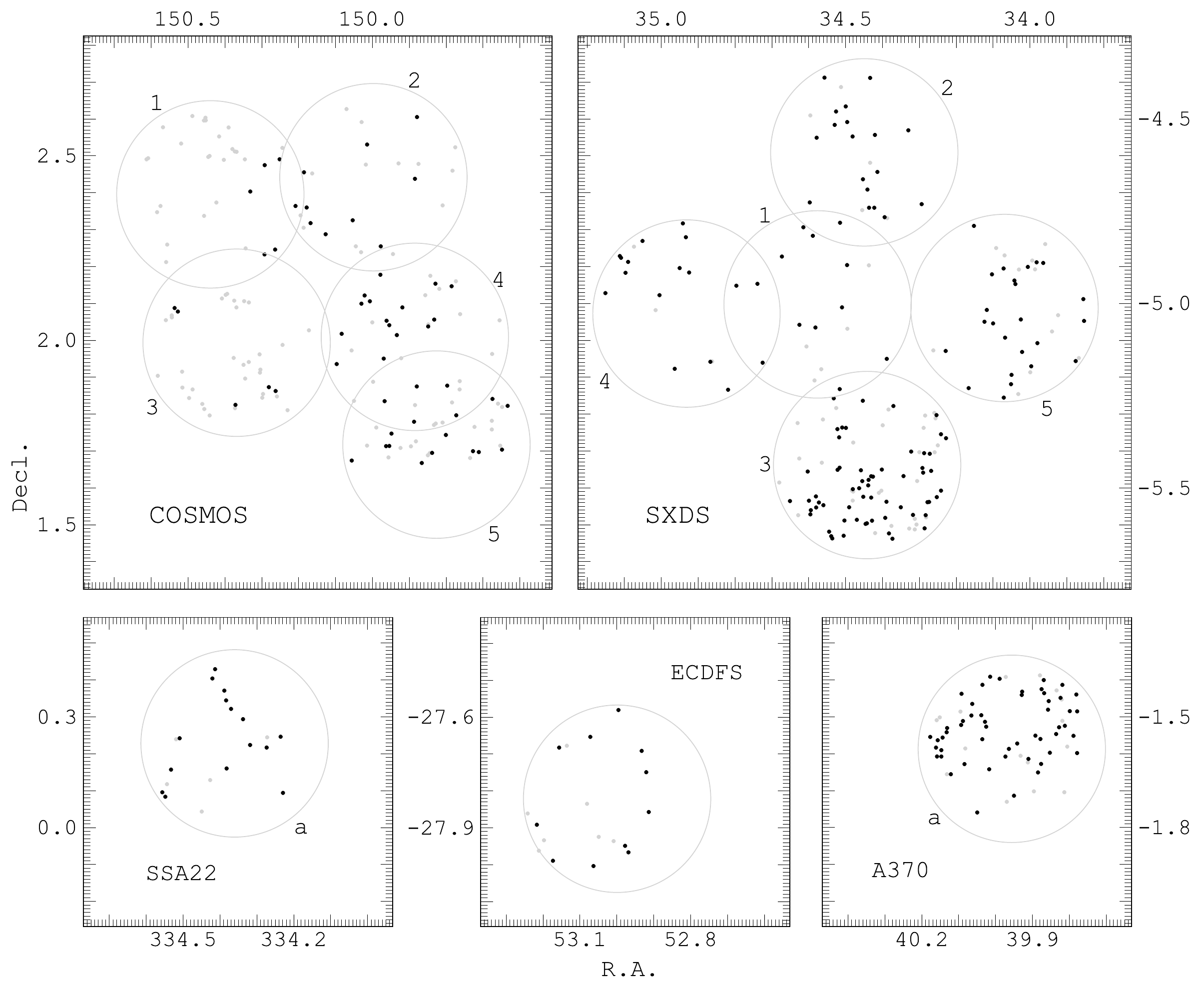}
\caption{The five deep fields observed by our M2FS survey. The big circles represent our M2FS pointings. All points inside the circles represent the $z\approx5.7$ LAE targets observed by our M2FS survey. The black points represent the spectroscopically confirmed LAEs at $z\approx5.7$ presented in this paper.
\label{fields}}
\end{figure*}

\subsection{260 spectroscopically confirmed LAEs at $z\approx5.7$}

Our final sample consists of 260 LAEs at $z\approx5.7$. We show 40 LAEs in Table \ref{sample1}. They have the highest S/N in the sample. Column 4 lists the spectroscopic redshifts measured from the \lya\ lines (their errors are smaller than 0.001; see Section 4 for details). Columns 5-7 show their photometry in $i$, $z'$, and NB816, respectively. Column 8 lists the \lya\ luminosities. Column 9 shows their identity numbers in our M2FS program. No. 10 (SXDS2-020) and No. 40 (SXDS3-016) are the two bright LAEs used in Figures 2 and 3. The whole table is provided on line. Figure \ref{glaeshsnr} shows the 1D and 2D spectra of the 40 LAEs in the sample. The whole sample is also provided on line. We can see that strong emission lines usually show asymmetric line shapes due to the IGM absorption and ISM kinematics. The 1D spectra in these figures are shown in arbitrary units for clarity. \lya\ line flux will be calculated using the narrowband and broadband photometry in Section 4.

Figure \ref{fields} illustrates the positions of the targets in the five fields, including the observed candidates (all points) and the confirmed LAEs (black points). The big circles represent the M2FS pointings. Despite the fact that the exposure time and depth of individual pointings are quite similar, the numbers of LAEs (Table \ref{fieldsinfo}) in these pointings are quite different, suggesting the existence of significant cosmic variance. Such cosmic variance was not due to selection bias during our spectroscopic observations, because it already exists in our photometrically selected candidates (see also Figure \ref{fields}). It has also been reported in previous studies \citep[e.g.,][]{ouchi08,hu10,kashikawa11,jiang18}. SXDS3 contains a giant protocluster at $z\approx5.7$, and thus has the largest number of LAEs. As we mentioned above, COSMOS1 and COSMOS3 have serious alignment problems during the observations, so they only have a few LAEs confirmed here. We will exclude these two pointings when we calculate the \lya\ LF in a following paper.

It is worth pointing out that the five well-studied fields have been previously used to search for high-redshift LAEs. For example, \citet{ouchi08} constructed a large photometric sample of $z\approx5.7$ LAEs in SXDS, and spectroscopically confirmed 17 of them. \citet{hu10} provided a spectroscopic sample of $z\approx5.7$ LAEs in several fields including A370a. We included these fields to crosscheck our target selection and sample completeness. As we already discussed in \citet{jiang17}, we recovered the above known LAEs in SXDS and A370a, suggesting a high completeness in our sample.

\section{\lya\ Spectral Properties}

In this section, we will measure the \lya\ spectral properties of the 260 LAEs in our sample. We will first measure their redshifts. We will then calculate their \lya\ line flux, rest-frame EW and UV continuum flux based on the secure redshifts and the NB816 and $z'$ band photometry. Next, we will also analyze the relation between the full width at half maximum (FWHM) of the \lya\ lines and \lya\ luminosities.

\subsection{Redshifts}

We use composite spectrum templates to calculate LAE (\lya) redshifts. For each LAE, we first estimate an initial redshift using the wavelength of the \lya\ line peak. We then refine this redshift by fitting the line using the composite template of the \lya\ profile from \citet{kashikawa11}. The central wavelength of the template \lya\ line is $\lambda_{\rm Ly\alpha}=1215.67$ \AA, and the line is scaled so that the peak value is 1 (arbitrary units). From this template, we generate a set of model spectra for a grid of peak value, line width, and redshift. The peak value, by scaling the composite line, is from 0.9 to 1.1 with a step size of 0.01. The line width, by shrinking and expanding the composite line, is from 0.5 to 2.0 times the original width with a step size of 0.1 (times the original width). The redshift value varies within the initial redshift $\pm0.002$ with a step size of 0.0001. Finally, we fit the \lya\ line of the LAE using the above model spectra and find the best fit. After we obtain the refined redshift for each LAE, we combine our spectra (weighted average) to produce a new template of the \lya\ line profile from our own sample (see section 5.1). We then repeat the above procedure a few times using our own template. The spectroscopic redshifts for 40 LAEs are shown in Table 3.

\begin{figure}[t]
\epsscale{1.15}
\plotone{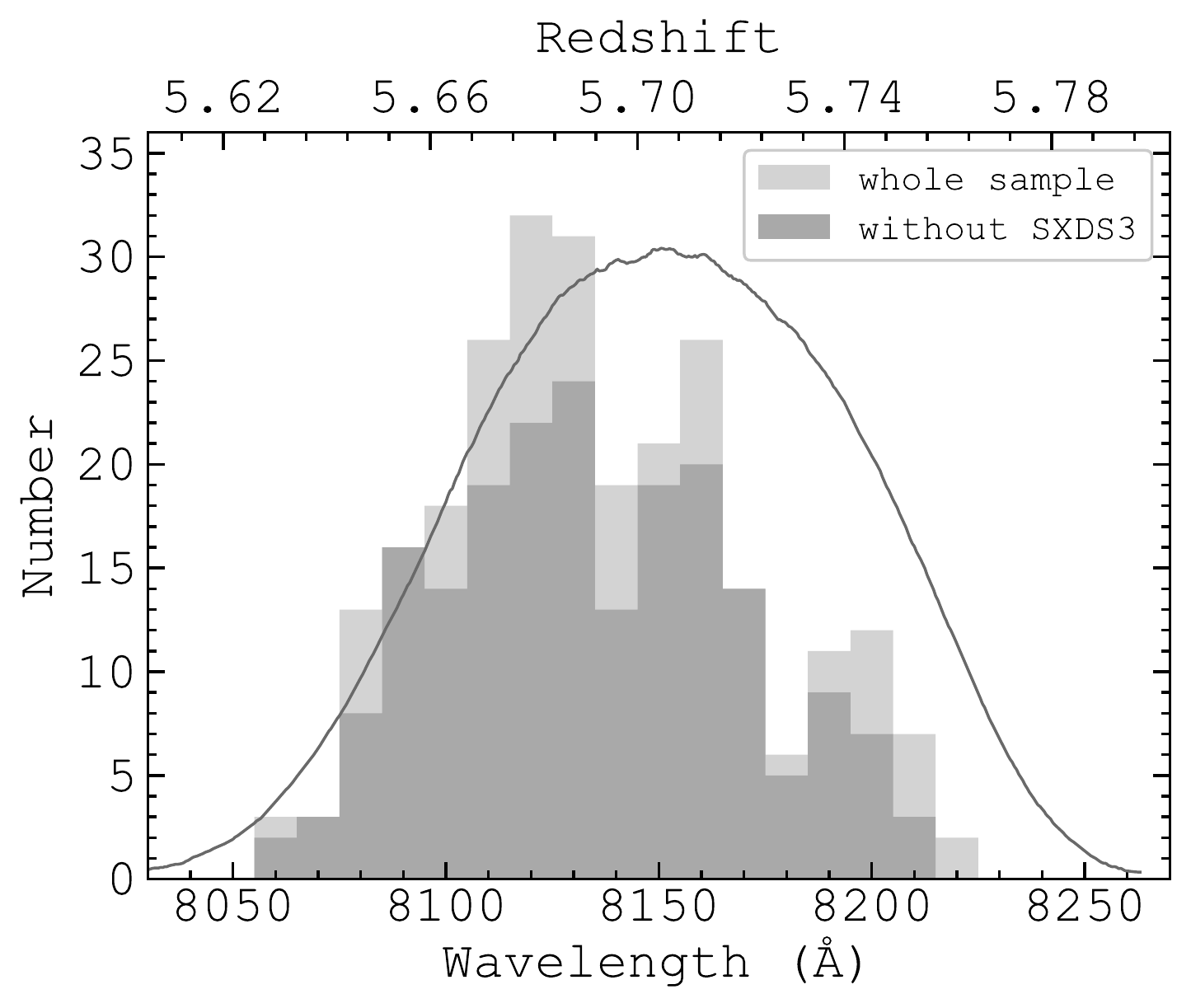}
\centering
\caption{Redshift distribution of the LAE sample. The light gray histogram represents all LAEs in our sample. The dark gray histogram represents the sample excluding LAEs in SXDS3 where there is a giant protocluster. The NB816 filter transmission curve is over-plotted and scaled for clarity. We see a clear offset of $\sim20$ \AA\ between the observed \lya\ wavelength distribution and the filter transmission curve.
\label{spechist}}
\end{figure}

Figure \ref{spechist} shows the redshift distribution of the LAEs in our sample (light gray histogram). The dark gray histogram represents the sample excluding LAEs in SXDS3 where there is a giant protocluster. We see an apparent mismatch between the redshift distribution and the NB816 filter transmission curve (black profile in the figure). We use a Gaussian profile to fit the redshift distribution of the dark gray histogram and compare the best fit to the filter transmission curve. The result indicates an offset of $\sim20$ \AA. This large offset is mainly due to the IGM absorption blueward of \lya\ in the high-redshift spectra. We will discuss this in Section 5.

\subsection{\lya\ Flux and Equivalent Width}

We use the narrowband (NB816) and broadband ($z'$) photometry to estimate the \lya\ line flux and UV continuum flux using a model spectrum. The model spectrum is the sum of a \lya\ emission with our template line profile and a power-law UV continuum with a slope $\beta$,
\begin{eqnarray}
   f_{\lambda} = f_{\rm Ly\alpha}\times{P}_{\rm Ly\alpha}(\lambda) + f_{\rm cont}\times\lambda^{\beta},
\end{eqnarray}
where $P_{\rm Ly\alpha}(\lambda)$ is the dimensionless line profile of our template that is redshifted to the observed frame for each individual LAEs, and $f_{\rm Ly\alpha}$ and $f_{\rm cont}$ in units of $\rm erg\ s^{-1}\ cm^{-2}$ \AA$^{-1}$ are scale factors of the \lya\ line flux and the UV continuum flux, respectively. We are not able to determine $\beta$ for individual LAEs, so we adopt an average $\beta=-2.3$ from a sample of spectroscopically confirmed LAEs at $z\ge5.7$ by \citet{jiang13a}. We then calculate $f_{\rm cont}$ from the $z'$ band photometry because the $z'$ filter does not cover \lya\ for $z\approx 5.7$ LAEs. For the LAEs that are not detected in the $z'$ band, we use $2\sigma$ detection upper limits. In this case, the (very weak) continuum flux has negligible impact on the measurement of the \lya\ flux below, because the narrowband photometry is completely dominated by the \lya\ flux.

After $f_{\rm cont}$ is determined, we using Equation 3 to calculate the \lya\ flux or  scale factor $f_{\rm Ly\alpha}$ by matching the model spectrum to the narrowband photometry. The IGM absorption is considered in the model spectrum. For simplicity, we assume that the flux blueward of \lya\ is completely absorbed. The \lya\ line shape has negligible impact, because the line width is much smaller than the narrowband filter width. After $f_{\rm Ly\alpha}$ is determined, we calculate UV luminosity $M_{1500}$, \lya\ luminosity, and \lya\ rest-frame EW. The measured \lya\ luminosities and EWs are not corrected for IGM absorption. EWs are given in the rest frame.

\begin{figure}[t]
\epsscale{1.15}
\centering
\plotone{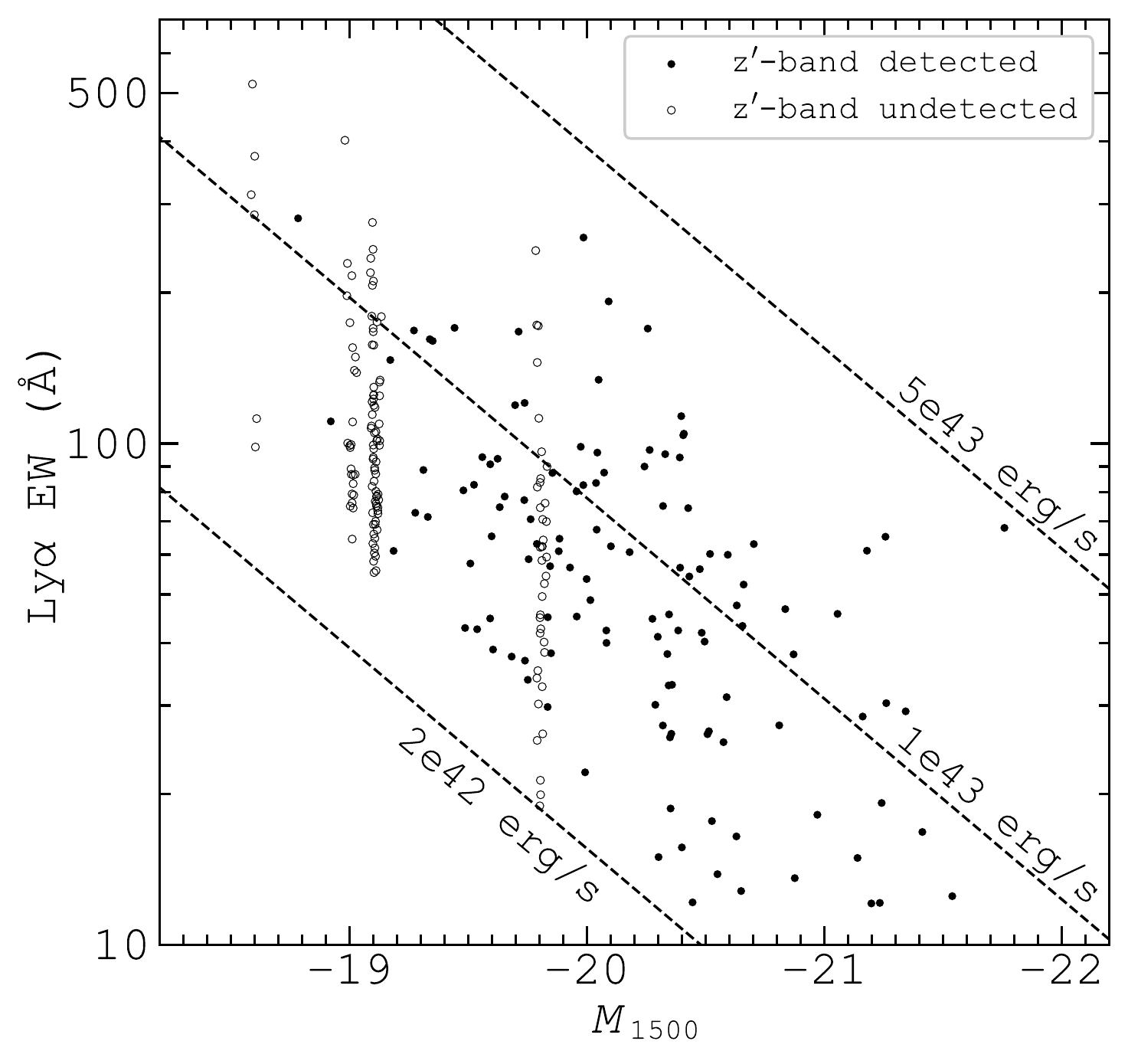}
\caption{\lya\ EW as a function of $M_{1500}$. The filled (open) circles represent the LAEs detected (undetected) in the $z'$ band. The diagonal dashed lines are defined by \lya\ luminosities. LAEs with lower UV luminosities tend to have higher \lya\ EWs, mostly due to selection effects.
\label{magew}}
\end{figure}

Figure \ref{magew} shows the \lya\ EWs as a function of $M_{1500}$. The filled circles represent the LAEs detected in the $z'$ band. The EWs of most LAEs range between 20 and $\sim300$ \AA\ and the median value is 62 \AA. The open circles represent the LAEs that are not detected in the $z'$ band. These LAEs potentially have larger EWs. Among them, five LAEs have EW $\gtrsim300$ \AA. When we include these LAEs, the median \lya\ EW value is 75 \AA, consistent with those given in \citet{kashikawa11} and \citet{jiang13a}. We will discuss extremely large \lya\ EWs in Section 5.

Figure \ref{magew} shows an apparent anti-correlation between \lya\ EW and $M_{1500}$, i.e., LAEs with lower UV luminosities tend to have larger EWs. This relation has been extensively discussed previously \citep[e.g.,][]{ouchi08, cowie10, cowie11a, jiang13a} and is mostly caused by selection effects. In Figure \ref{magew}, the three dashed lines indicate \lya\ luminosities of $2\times10^{42}$, $1\times10^{43}$, and $5\times10^{43}$ erg s$^{-1}$, respectively. The first \lya\ luminosity roughly corresponds to the flux limit of our survey. In a narrowband flux-limited survey, LAEs with weak continuum emission and small \lya\ EWs will not be selected. On the other hand, LAEs with large EWs and high UV luminosities should be easily included. However, there are no LAEs with $M_{1500}<-20.5$ mag and EW $>80$ \AA\ in our sample, as shown in Figure 8. The lack of such LAEs in our sample and in previous studies indicates that these galaxies are extremely rare.

\begin{figure*}
\centering
\includegraphics[angle=0, width=1\textwidth]{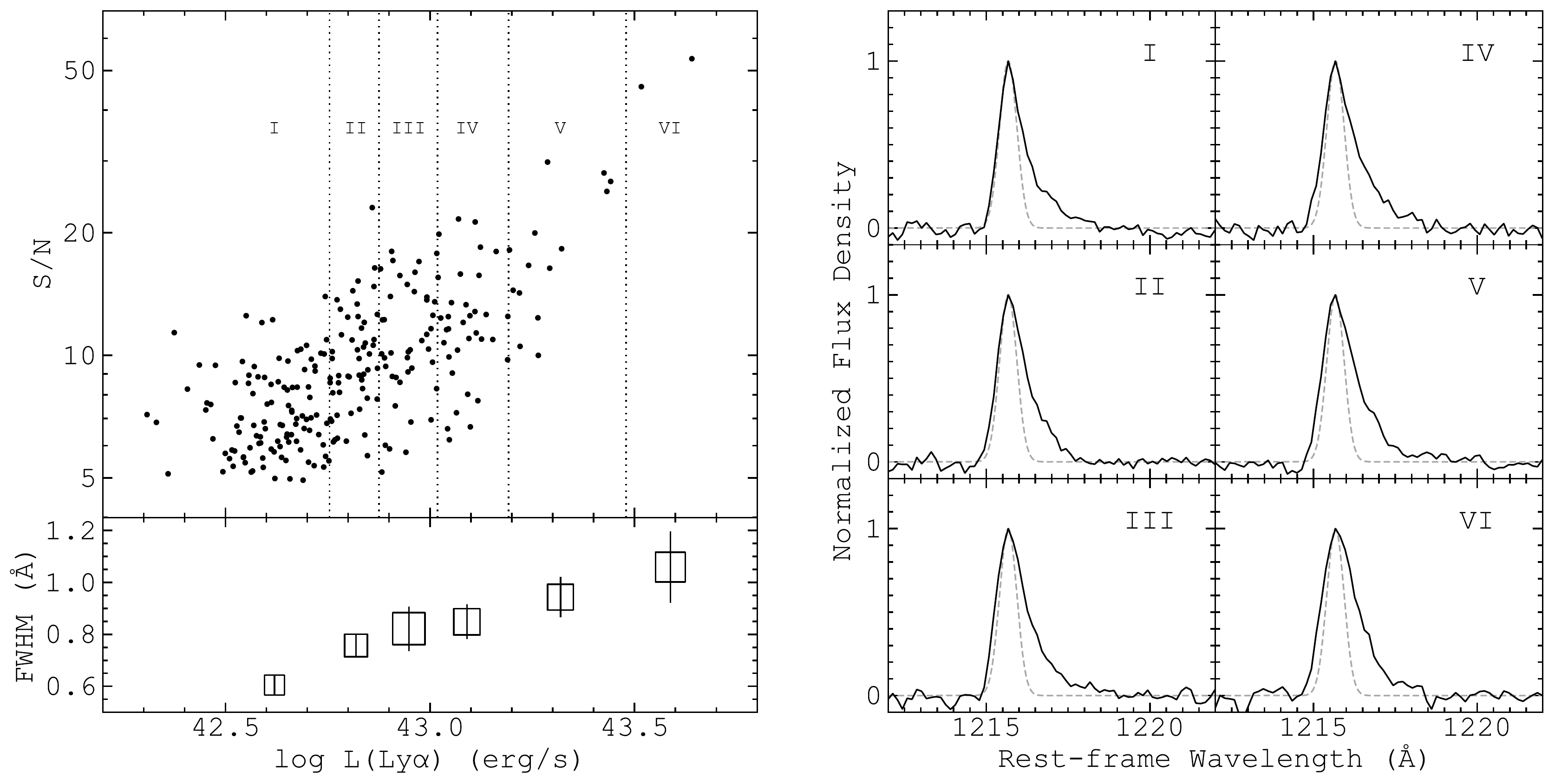}
\caption{{\it Left:} In the upper panel, we divide our sample into six subsamples from I to VI by the five dotted vertical lines. We then make a combined spectrum for each subsample. In the lower panel, we show the intrinsic line widths after the correction of the instrument broadening. It clearly shows that the \lya\ line width increases towards higher luminosities.
{\it Right:} The combined spectra for the six subsamples in the left panel. The spectra have been normalized so that the peak flux density is 1. The dashed lines represent the instrument resolution (assuming a Gaussian profile). 
\label{lyastack}}
\end{figure*}

\subsection{\lya\ Line Profile}

The \lya\ emission line shape can be used to infer the distribution and kinematics of gas and the energy power of radiating sources due to the propagation and attenuation of \lya\ photons \citep{dijkstra14}. Here we investigate the relation between the \lya\ FWHM and luminosity from our sample. Most of the individual LAE spectra do not have sufficient S/N for reliable shape measurements, so we divide our sample into six subsamples based on the \lya\ luminosities. We then build a combined \lya\ spectrum for each subsample and measure FWHMs for six combined spectra.

In the left of Figure \ref{lyastack}, we show how the sample is divided into six subsamples I to VI based on the \lya\ luminosities. We have excluded 5 lowest-redshift LAEs whose \lya\ lines at $\le8070$ \AA\ are severely affected by relatively strong skylines. We have also excluded 5 LAEs with the lowest S/N. The sample is divided so that the final six combined spectra have similar S/N $\sim75$. The separation luminosities are log $L$(\lya) = 42.75, 42.88, 43.02, 43.19, 43.48 ($L$ is in units of erg s$^{-1}$). Note that there are only two LAEs in the brightest subsample. For each subsample, we co-add individual spectra to make a weighted average spectrum in the rest frame. The results are shown in the right panel of Figure \ref{lyastack}.

Figure \ref{lyastack} clearly suggests that the \lya\ line width increases towards higher luminosities. The luminosities are the weighted average luminosities in individual subsamples. We estimate the intrinsic \lya\ line width ${\rm FWHM}$ from the observed line width ${\rm FWHM_{obs}}$ and instrument resolution ${\rm FWHM_{ins}}$ using ${\rm FWHM^2} = {\rm FWHM^2_{obs}}-{\rm FWHM^2_{ins}}$. All FWHM values are converted to the rest-frame values. We estimate the instrument resolution by measuring strong sky emission lines near the $z\approx5.7$ OH-dark window. The resultant resolving power is $R\approx1954$. A Gaussian profile with ${\rm FWHM_{ins}}$ is shown as the light dashed lines in the right panel of Figure \ref{lyastack}. 
In the lower-left panel, the open squares represent the calculated FWHMs. For each square, its size indicates the measurement error from the combined spectrum in this subsample, and the vertical error bar indicates the standard deviations of FWHMs from a number of combined spectra. These combined spectra are generated from random groups of individual spectra in this subsample.

Given the high-quality spectra, the trend is significant, indicating that the \lya\ line width increases from 0.60 $\pm$ 0.04 to 1.06 $\pm$ 0.14 \AA. The second most luminous subsample has an average FWHM of 233 $\pm$ 19\ km s$^{-1}$, consistent with those of LAEs with similar luminosities at $z\approx5.7$ in \citet{matthee17}. The line widths suggest that the contribution from AGN activity is negligible on average in our LAEs \citep[e.g.,][]{matthee17}. The luminosity-dependent \lya\ FWHM has been observed in previous work \citep[e.g.,][]{hu10, matthee17, songaila18}. By fitting a power-law relation of FWHM = $AL^{\alpha}$, we obtain $\alpha=0.25\pm0.04$, which is consistent with the result in \citet{hu10}. Such a trend has been predicted theoretically \citep[e.g.,][]{sadoun19}. On average, more luminous LAEs reside in more massive halos with higher gas velocities and higher neutral hydrogen column densities. Both of them would increase the \lya\ line width through radiative transfer, leading to the luminosity-dependent FWHM.

\section{Discussion}

\subsection{Composite Spectra}

High-ionization metal lines in the rest-frame UV band are useful tools to study star-forming galaxies. However, it is difficult to detect UV emission lines except \lya\ in individual spectra of high-redshift galaxies. Our sample includes 260 LAEs, and we did not detect UV emission lines near \lya\ in the individual spectra. Composite spectra are often constructed to study line features that are too weak to be detected in individual spectra \citep[e.g.,][]{shapley03, jones12, zheng16}. In this section, we combine our spectra and try to detect UV emission lines in the combined spectra. Our spectra do not cover most of the commonly found lines such as C\,{\sc iv} $\lambda 1549$, He\,{\sc ii} $\lambda 1640$, or C\,{\sc iii}] $\lambda 1909$. The only exception is the \nv\ $\lambda\lambda1239,1243$ doublet, which is close to \lya.

\begin{figure}[t]
\epsscale{1.15}
\centering
\plotone{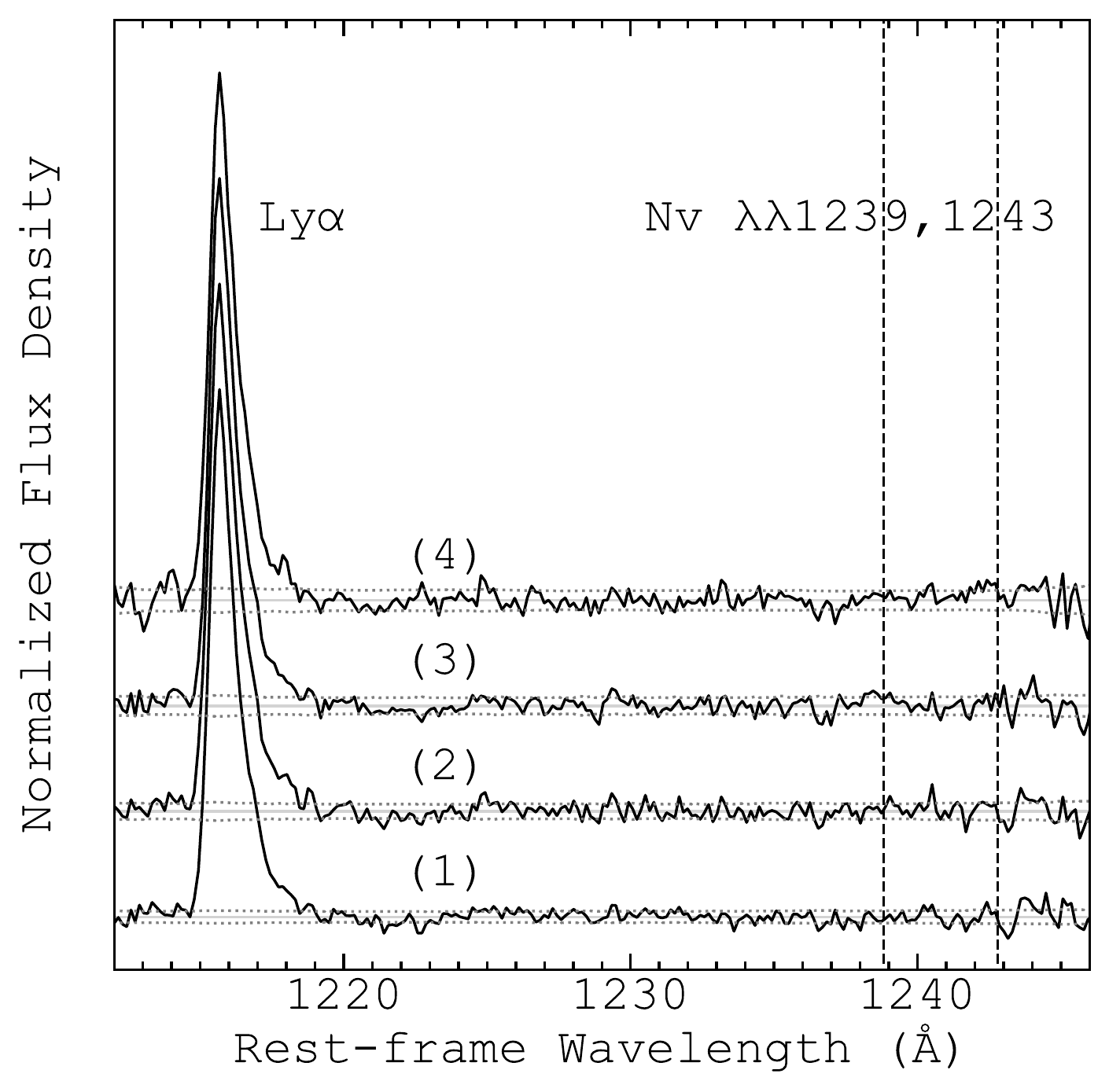}
\caption{The normalized composite spectra of (1) all LAEs; (2) 66 LAEs with log $L$(\lya) $>43$ ($L$ is in units of erg s$^{-1}$); (3) 70 LAEs with EW $>100$ \AA; and (4) 41 $z<5.662$ LAEs whose \nv\ emission is not affected by strong skylines. The spectra 2, 3, and 4 have been shifted for clarity. The gray dotted lines indicate the $1\sigma$ error regions. The vertical dashed lines indicate the wavelengths of the expected \nv\ $\lambda\lambda1239,1243$ doublet. The \nv\ emission is not detected in these spectra.
\label{specomb}}
\end{figure}

We combine all spectra and spectra in three subsamples based on \lya\ luminosity, EW, and redshift, respectively. The individual spectra are normalized and converted to the rest frame. Then they are combined (weighted average) with a rejection of $>5\sigma$ outliers. The results are shown in Figure \ref{specomb}. The four spectra represent the composite spectra of (1) all LAEs; (2) 66 LAEs with log $L$(\lya) $>43$ ($L$ is in units of erg s$^{-1}$); (3) 70 LAEs with EW $>100$ \AA; and (4) the 41 lowest-redshift ($z<5.662$) LAEs whose \nv\ emission is not affected by strong skylines (no outlier rejection in the spectral combination). The two vertical dashed lines indicate the expected positions of the \nv\ doublet. No \nv\ emission is detected.

The \nv\ emission has a very high ionization potential and has been very rarely found in star-forming galaxies. It can be used to search for AGN activity in luminous LAEs \citep{sobral18b}. Unlike LAEs in the local universe, the AGN fraction at high redshift is small \citep{ouchi08, zheng10}. 
Based on the non-detection above, the $3\sigma$ upper limit of the \nv\ flux estimated from the local noise level, is about 1.1\% of the \lya\ flux. Recently, \citet{guo20} combined $\sim150$ spectra of LAEs at $z\approx3.1$ and detected the \nv\ $\lambda1239$ emission line and the C\,{\sc iv} $\lambda\lambda1548,1551$ doublet lines at $\sim4\sigma$ level. Their flux ratio of \nv/\lya\ is about 0.7\%, smaller than the ratio in our combined spectrum. Therefore, the non-detection of \nv\ in Figure \ref{specomb} is reasonable if we assume that our LAEs are the higher-redshift counterparts of the $z\approx3.1$ LAEs. This also indicates that the AGN contribution is negligible in our sample.

\subsection{\lya\ EW}

\begin{figure}
\epsscale{1.15}
\centering
\plotone{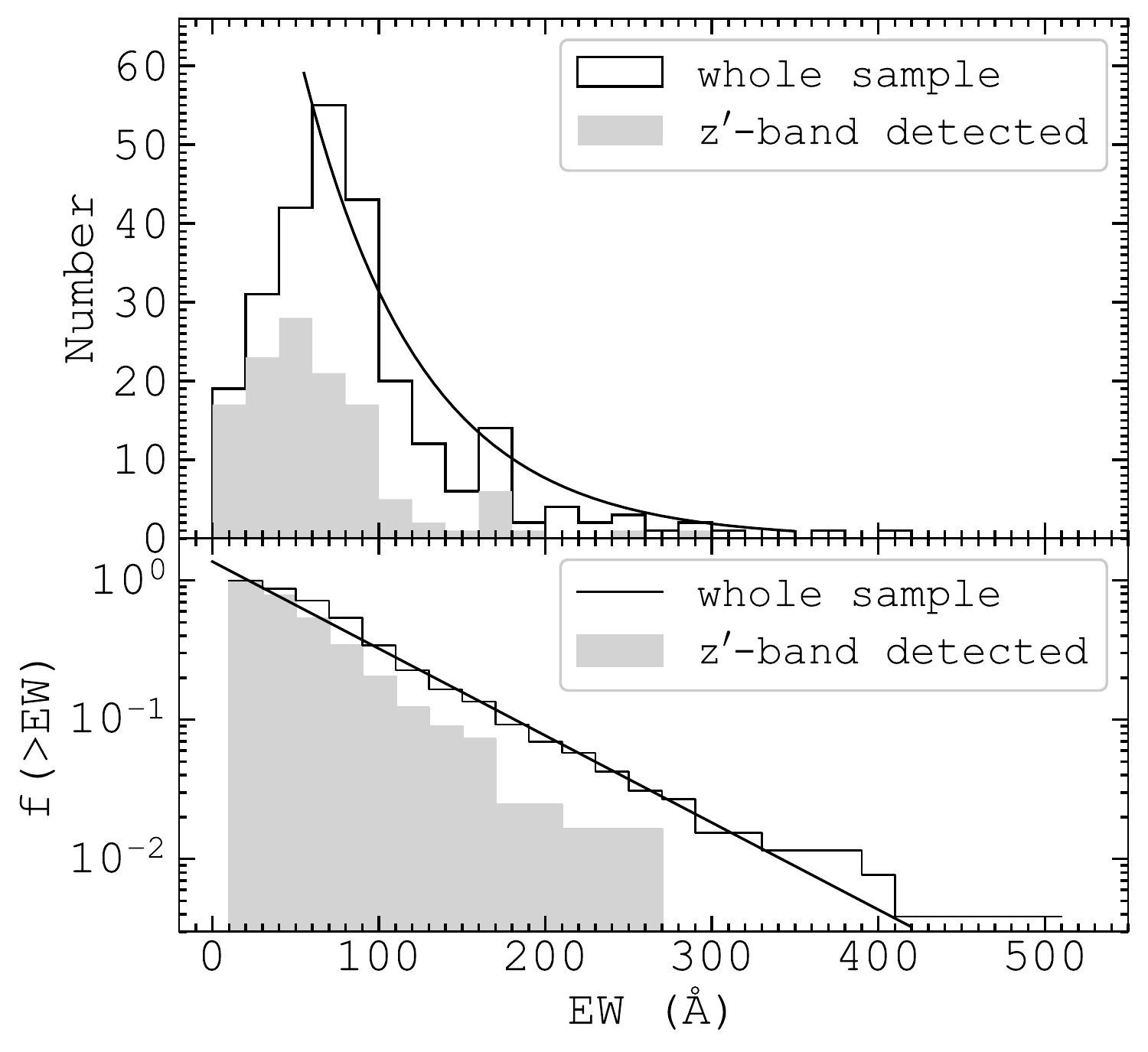}
\caption{{\it Upper panel:} \lya\ EW distribution of the LAE sample. The filled histogram represents the LAEs detected in $z'$ and the solid line histogram represents the whole sample. The curve is the exponential function fitted to the histogram of the whole sample with a lower limit of 50 \AA. 
{\it Lower panel:} Comparison of the cumulative \lya\ EW fraction distributions between the two samples shown in the upper panel. See the text for details.
\label{ewDist}}
\end{figure}

We show the \lya\ EW distribution of the sample in the upper panel of Figure \ref{ewDist}. The filled histogram represents the LAE sample detected in the $z'$ band. The solid line histogram represents the whole sample, including those undetected in $z'$. For LAEs undetected in $z'$, the $2\sigma$ upper limits are used for the $z'$-band photometry. The EW distribution can be described by an exponential form $dN/d{\rm EW} \propto {\rm exp}(-{\rm EW}/W_0)$ with a characteristic $e$-folding EW scale $W_0$ \citep[e.g.,][]{shapley03, cowie10}. We use the exponential function to fit the histogram within $\rm EW=50$ and 300 \AA. The lower panel shows the  cumulative EW fraction distribution $\rm f(>EW) = exp(-EW/W_0)$. We obtain a scale length $W_0 = 70\ \pm\ 2$ \AA\ from our sample. This value is underestimated because a large fraction of LAEs were not detected in $z'$. 
Previous studies have shown that the \lya\ EW slowly increases from low redshift to $z\sim6$ \citep[e.g.,][]{wold14, wold17, zheng14, hashimoto17a}, and then declines towards higher redshift due to its resonant scattering by neutral hydrogen in the IGM \citep[e.g.,][]{jung18, mason18a}. So, the \lya\ EW distribution of high-redshift LAEs ($z\gtrsim6$) can be used to probe the reionization history \citep[e.g.,][]{jung18, mason18a}. Our result of the \lya\ EW distribution is generally consistent with previous studies at $z\lesssim6$ \citep[e.g.,][]{zheng14, jung18}. 

Figure \ref{ewDist} also shows that $\sim7$\% of our LAEs have \lya\ EWs greater than 200 \AA, including at least seven LAEs with $\rm EW \gtrsim300$ \AA. These LAEs with such high \lya\ EWs tend to hold stellar populations with very low metallicity and young stellar age \citep{cf93}. As shown in Figure \ref{magew}, these LAEs have relatively weak UV continua. Figure \ref{magew} also shows that there is a deficit of large-EW LAEs with bright UV continua. In high-redshift star-forming galaxies, \lya\ photons are more absorbed than UV continuum photons due to the complex \lya\ radiative transfer in the ISM \citep{dijkstra14}. Based on the positive correlation between \lya\ escape fraction and EW \citep{sobral19}, bright-continuum LAEs tend to have smaller \lya\ escape fraction or stronger dust attenuation. This relation is consistent with the picture that bright-continuum LAEs hold intense star formation which boosts the metal/dust enrichment.

\subsection{Redshift Distribution}

\begin{figure}[t]
\epsscale{1.15}
\centering
\plotone{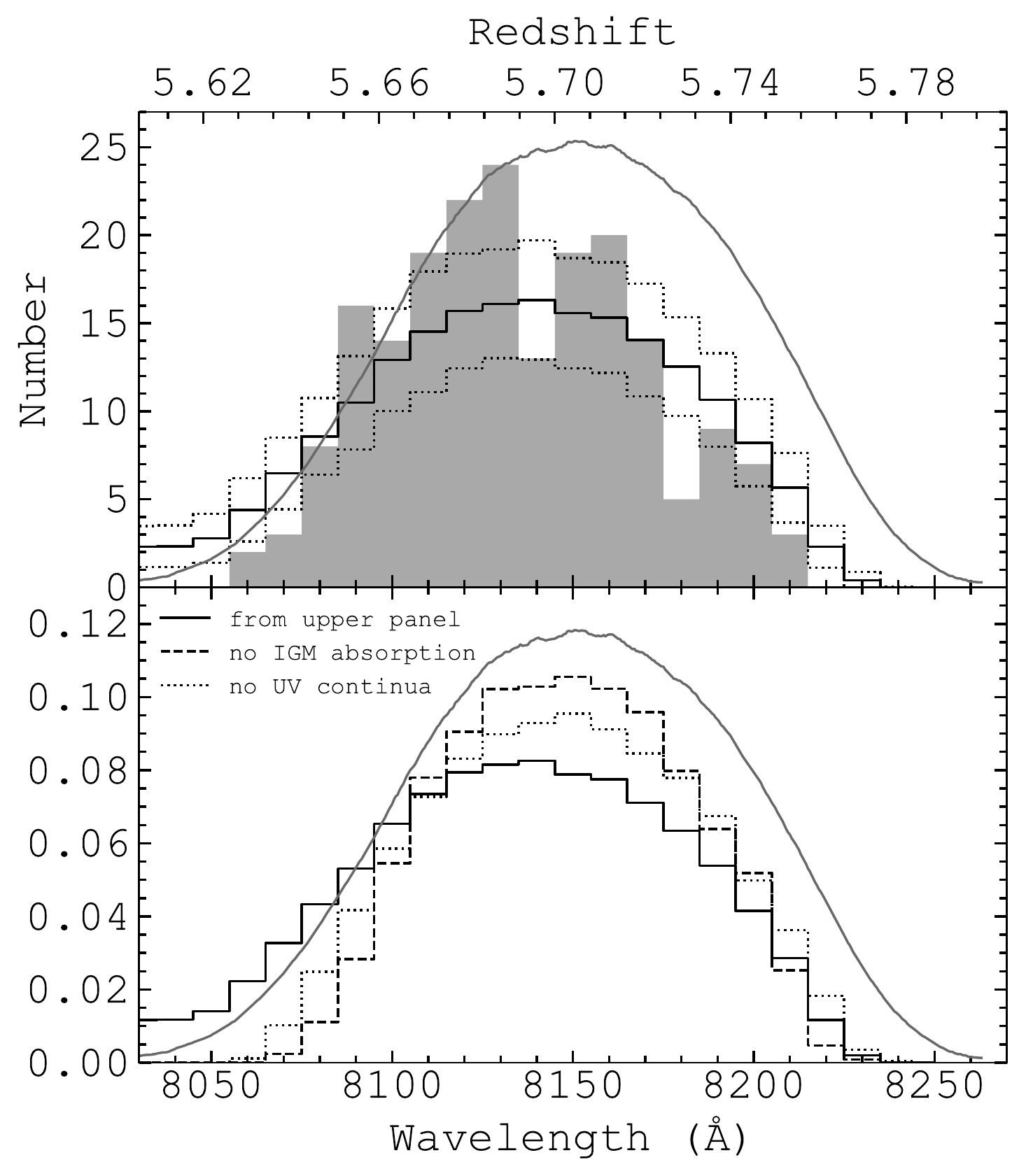}
\caption{Redshift distribution of mock LAEs. The curves represent the NB816 filter transmission curve. {\it Upper panel:} Distribution of the mock LAE samples compared with the observed distribution. The gray histogram represents the observed distribution excluding the LAEs in SXDS3. The wavelength bins are the same as those shown Figure \ref{spechist}. The solid step histogram represents the average distribution of the 250 mock samples and the two dotted step histograms represent its $1\sigma$ error range.  {\it Lower panel:} Normalized distributions of the mock LAEs. The solid step histogram is the same as that in the upper panel. The dashed histogram represents a case in which the mock LAEs are not corrected for the IGM absorption. The dotted histogram represents a case in which the mock LAEs do not have UV continua. 
\label{zDist}}  
\end{figure}

In Figure \ref{spechist}, we show a clear mismatch between the redshift distribution of the sample and the NB816 filter transmission curve. In this section we will use a simulation to show that this mismatch is mainly caused by the IGM absorption of the continuum emission blueward of \lya\ in high-redshift spectra. In this simulation, we build samples of mock LAEs at $z\approx5.7$ and apply the following selection criteria, 
\begin{eqnarray}
   i-{\rm NB816}>1.0\ {\rm and}\ {\rm NB816}<26.0.
   \label{msc}
\end{eqnarray}
In the simulation, we consider \lya\ LF, EW distribution, broadband and narrowband filter transmission. Each mock LAE is initially assigned values for three quantities: redshift, \lya\ luminosity, and \lya\ EW. The UV continuum is calculated from the \lya\ luminosity and EW, assuming a constant UV slope $\beta=-2.3$. We then implement the IGM absorption blueward of \lya\ in the spectrum. The broadband and narrowband magnitudes are calculated from the redshifted mock LAE spectrum and the response curves of the $i$ and NB816 filters.

The redshift value is chosen to ensure that the \lya\ line is in the wavelength range of $8025 - 8255$ \AA\ (the NB816 bandpass range). The \lya\ luminosity is generated in the logarithmic range of $42.3 - 43.7$ based on the Schechter LF \citep{schechter76}, 
\begin{eqnarray}
   \phi({\rm log}L) = {\rm ln10}\ \phi^*\ (\frac{L}{L^*})^{\alpha+1}\ {\rm exp}(-\frac{L}{L^*}),
   \label{lyalf}
\end{eqnarray}
where $\phi^*=2.5\times10^{-4}\ {\rm Mpc}^{-3}$, ${\rm log}\ L^*=43.0$, and $\alpha=-1.5$ \citep{kashikawa11}. 
To generate EW values, we use an exponential form, 
\begin{eqnarray}
  dN/d{\rm EW} \propto {\rm exp}(-{\rm EW}/W_0),
  \label{ewdist}
\end{eqnarray}
where the EW scale is assumed to be $W_0=100$ \AA.

We generate 800 LAEs for one mock sample. These LAEs follow Equations \ref{lyalf} and \ref{ewdist}. We create 250 such samples independently, resulting in a total of 200,000 mock LAEs. In each sample, slightly more than 250 mock LAEs satisfy the selection criteria (\ref{msc}). We further assume that these LAEs would be securely identified in our spectra, so the number of the LAEs in each mock sample is similar to that in our real sample. These LAEs are assigned into the same wavelength or redshift bins shown in Figure \ref{spechist}. In Figure \ref{zDist}, the upper panel shows the average distribution of the 250 mock samples (solid step histogram) and its $1\sigma$ error (dotted step histograms). They are slightly scaled to match the gray histogram that represents our real sample without the LAEs in SXDS3. By repeating the same procedure with different $\beta$ values, we find that the assumption of the UV slope has negligible impact on our results.

Figure \ref{zDist} shows that the redshift distribution of the mock galaxies does not have a Gaussian shape. Its overall shape is well consistent with the redshift distribution of our LAE sample. We further demonstrate using two more simulations that the asymmetric shape of the redshift distribution is mainly due the the IGM absorption of the continuum emission blueward of \lya. In the first simulation, we do not correct for the IGM absorption, and the result is shown as the dashed histogram in the lower panel of Figure \ref{zDist}. In the second simulation, we assume that mock LAEs do not have continuum emission (just \lya\ emission), and the result is shown as the dotted histogram. The two histograms are nearly symmetric around the center of the NB816 filter. This means that we can use the IGM absorption to explain the offset between the LAE redshift distribution and the NB816 filter curve. We emphasize that it is mainly caused by the IGM absorption of the continuum emission, not the asymmetric \lya\ line emission.

\section{Summary}

We have presented a sample of 260 LAEs at $z\approx5.7$ in five well-studied fields, including SXDS, A370, ECDFS, COSMOS, and SSA22. It is by far the largest sample of spectroscopically confirmed LAEs at this redshift. The candidates were selected from the narrowband NB816 photometry and broadband photometry. The spectroscopic observations were carried out using M2FS on the Magellan Clay telescope. The whole sample was covered by 13 M2FS pointings with a total sky area of about 2 deg$^2$. The total on-source integration time was $\gtrsim5$ hrs per pointing. We identified LAEs based on the 1D and 2D M2FS spectra.

We have measured the \lya\ spectral properties of our LAEs. Assuming reasonable UV slopes, we used the NB816 and $z'$ band photometric data and the secure redshifts to derive \lya\ line flux, UV continuum flux, and \lya\ EW. We found that the EWs in our sample are mostly between 20 and 300 \AA, and these LAEs span a \lya\ luminosity range of $\sim 2\times10^{42} - 5\times10^{43}$ erg s$^{-1}$, including some of the most luminous galaxies known at $z \ge 5.7$. We also measured the FWHMs of the stacked \lya\ lines in different \lya\ luminosity bins. We found that the line width, after corrected for instrument broadening, clearly increases towards higher \lya\ luminosities.

Based on the narrow \lya\ line widths and the non-detection of \nv\ in the composite spectra, the AGN contribution is negligible in our sample. We have measured the LAE redshifts by fitting a composite \lya\ line template to the individual 1D lines. We discovered a large offset of $\sim20$ \AA\ between the observed \lya\ wavelength distribution and the NB816 filter transmission curve. Using the simulations, we explained that it is due to the IGM absorption of continua blueward of \lya\ in the high-redshift spectra.

\acknowledgments
We acknowledge support from the National Science Foundation of China (11721303, 11890693, 11991052), the National Key R\&D Program of China (2016YFA0400702, 2016YFA0400703), and the Chinese Academy of Sciences (CAS) through a China-Chile Joint Research Fund \#1503 administered by the CAS South America Center for Astronomy. This research uses data obtained through the Telescope Access Program (TAP), which has been funded by the National Astronomical Observatories of China (the Strategic Priority Research Program ``The Emergence of Cosmological Structures" Grant No. XDB09000000), and the Special Fund for Astronomy from the Ministry of Finance. We thank Z. Zheng for helpful discussion. This paper includes data gathered with the 6.5 meter Magellan Telescopes located at Las Campanas Observatory, Chile.

\facilities{Magellan: Clay (M2FS)}


\begin{figure*}
\centering
\includegraphics[angle=0, width=0.9\textwidth]{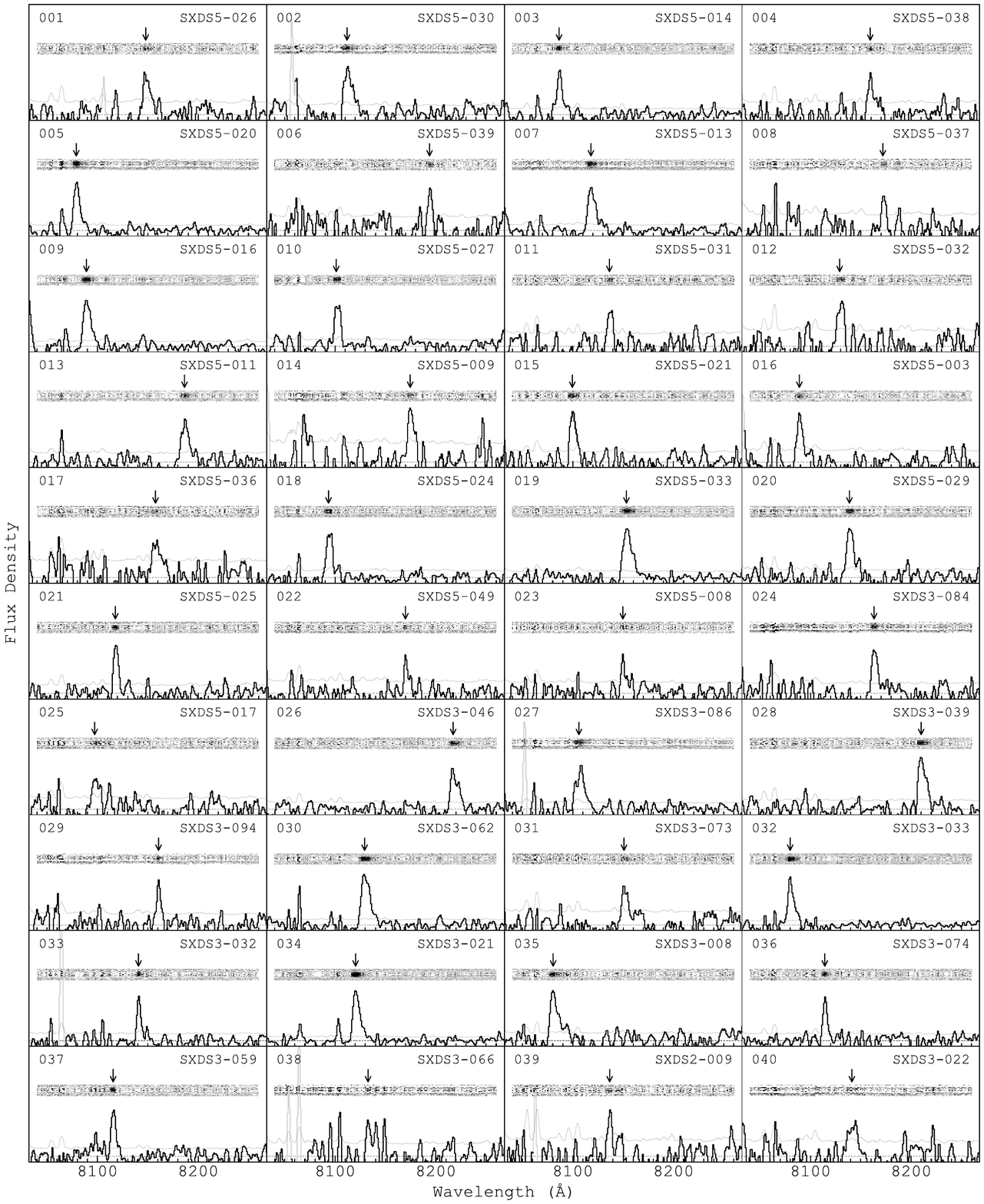}
\end{figure*}

\begin{figure*}
\centering
\includegraphics[angle=0, width=0.9\textwidth]{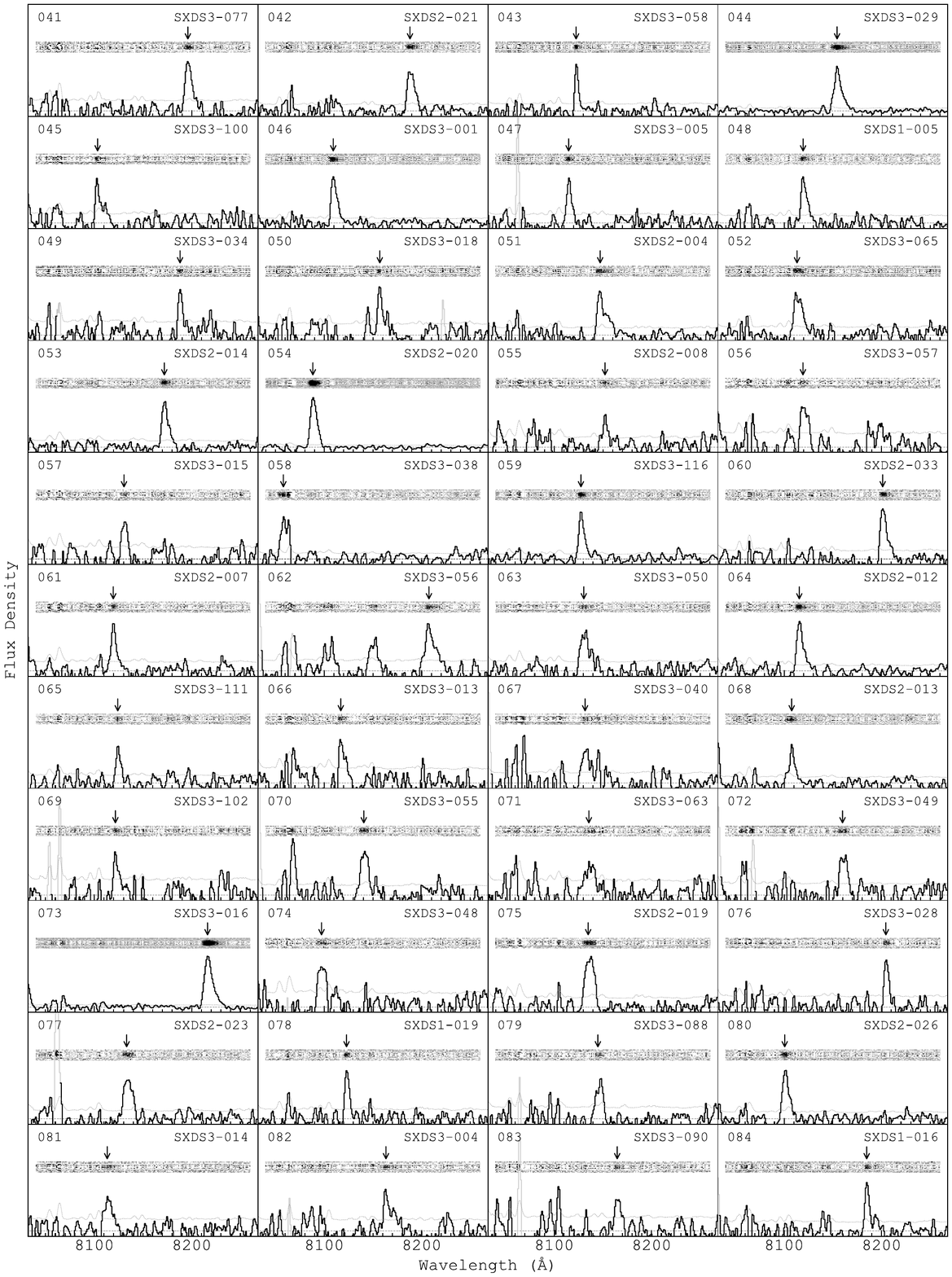}
\end{figure*}

\begin{figure*}
\centering
\includegraphics[angle=0, width=0.9\textwidth]{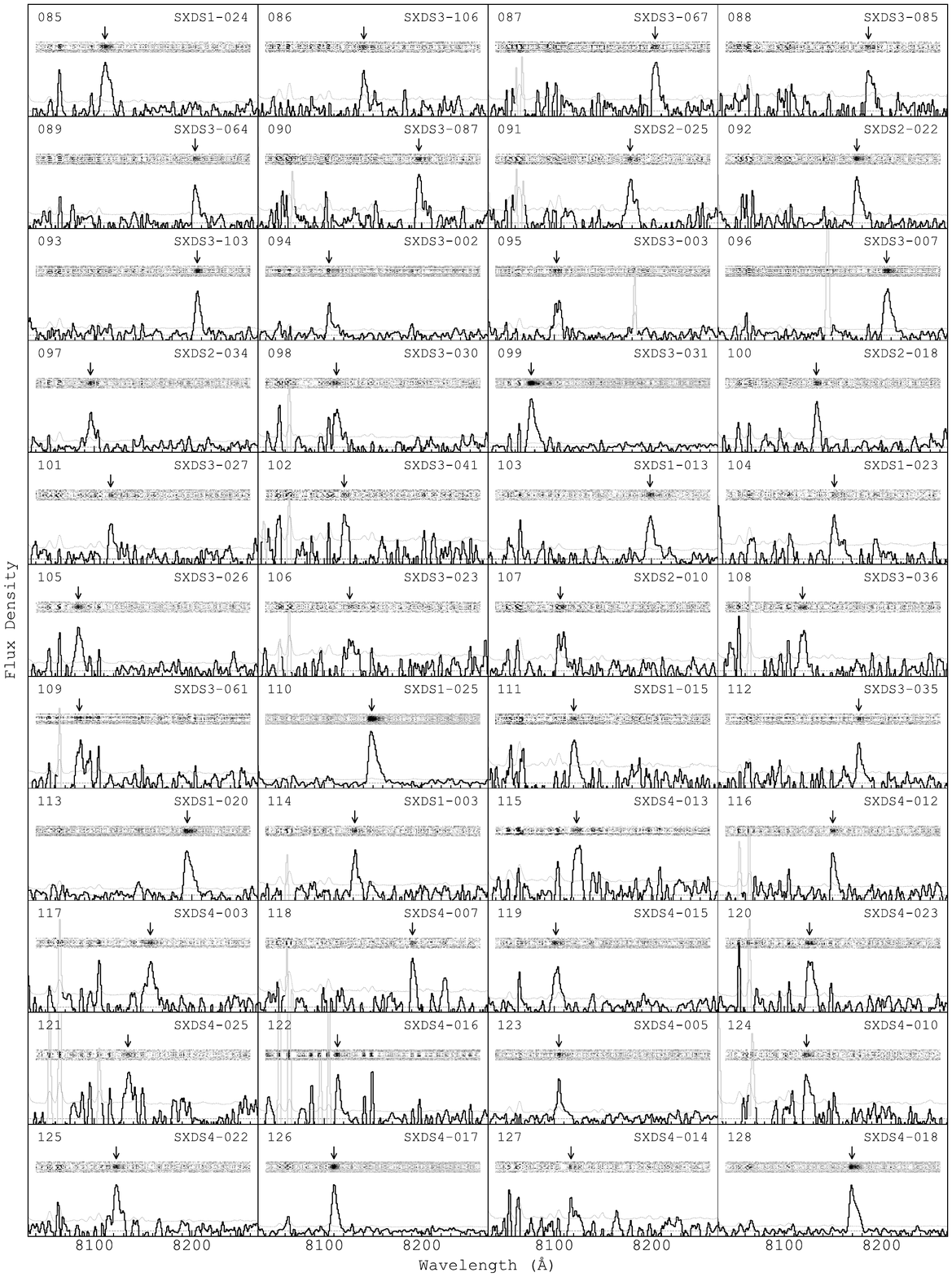}
\end{figure*}

\begin{figure*}
\centering
\includegraphics[angle=0, width=0.9\textwidth]{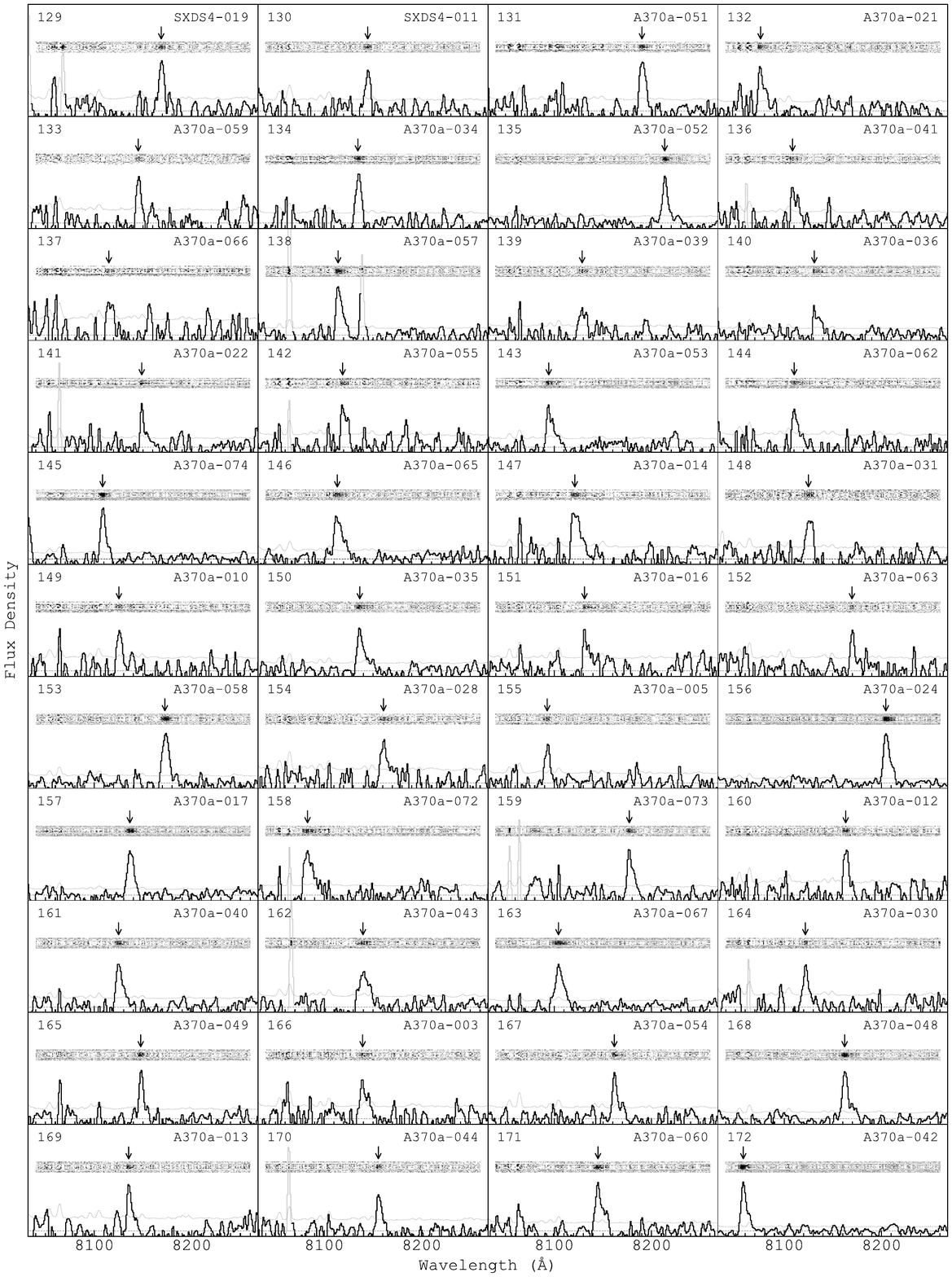}
\end{figure*}

\begin{figure*}
\centering
\includegraphics[angle=0, width=0.9\textwidth]{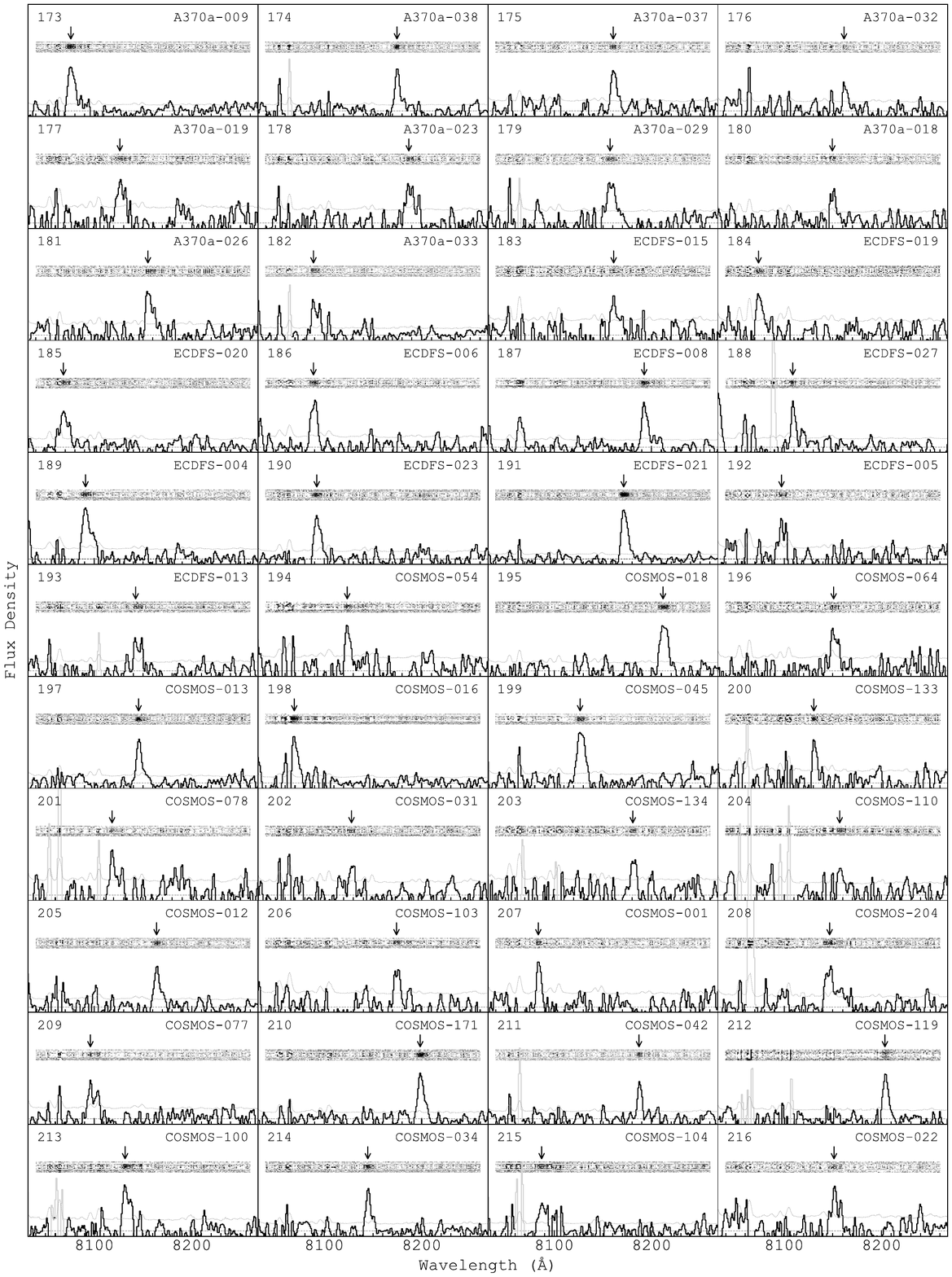}
\end{figure*}

\begin{figure*}
\centering
\includegraphics[angle=0, width=0.9\textwidth]{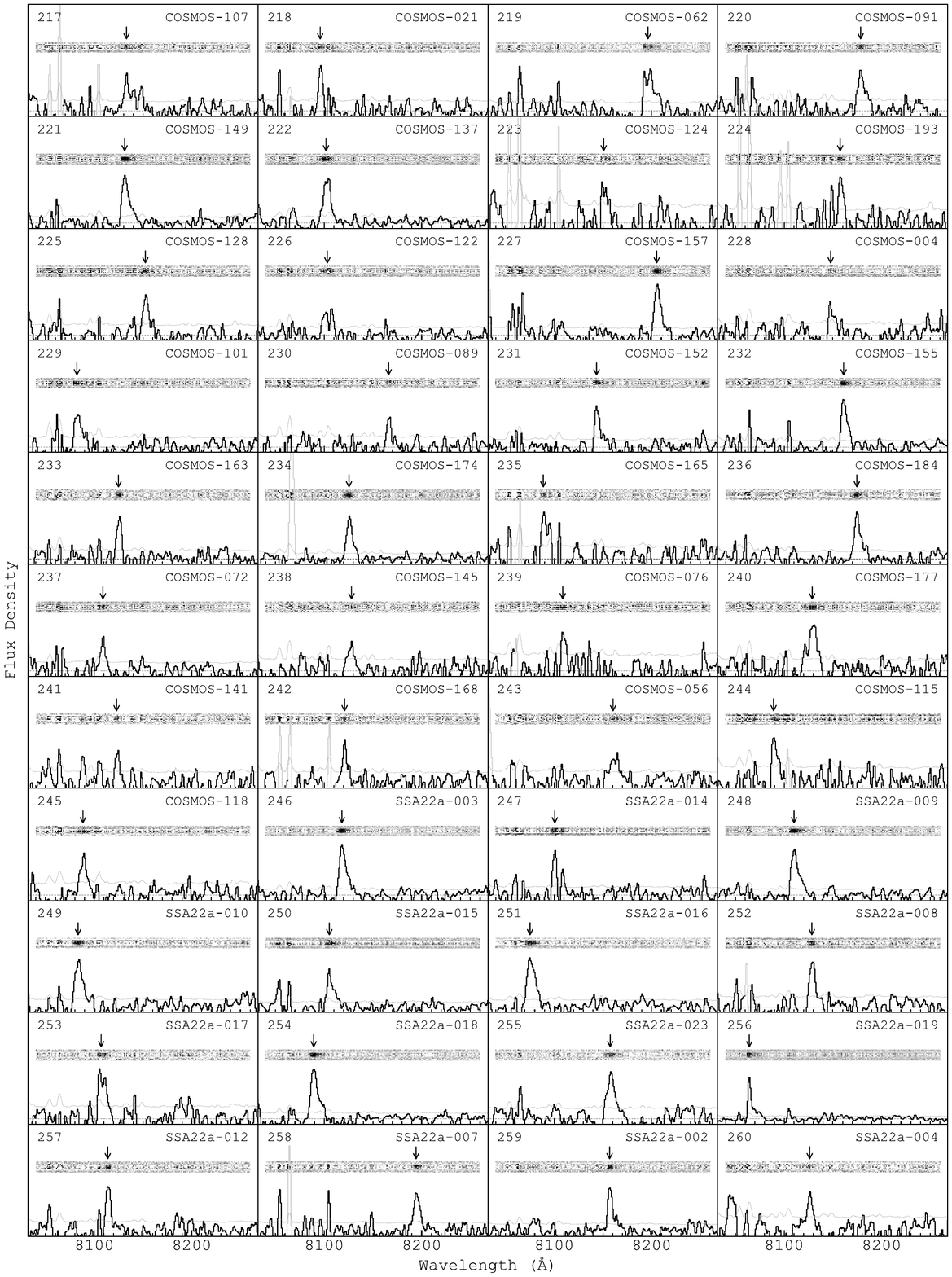}
\end{figure*}

\end{document}